\documentclass[aps,preprintnumbers,floatfix,article,amsmath,amssymb,floatfix,10pt,prd,superscriptaddress,nofootinbib]{revtex4}
\usepackage{bm}
\usepackage{amsfonts}
\usepackage{latexsym}
\usepackage[latin1]{inputenc}
\usepackage{graphicx}
\usepackage{amsmath}
\usepackage{palatino}
\usepackage{mathpazo}
\usepackage{textcomp}
\usepackage{float}
\usepackage{booktabs}
\usepackage{dcolumn}
\usepackage{ragged2e}
\usepackage{hyperref}
\hypersetup{colorlinks,citecolor=blue}
\hypersetup{colorlinks=true,linkcolor=red,filecolor=magenta,    urlcolor=cyan}
\usepackage{amsmath}
\usepackage{xcolor}
\usepackage{epsfig}
\usepackage{caption}
\usepackage{subcaption}
\usepackage{commath}
\def\jnl@style{\it}
\def\aaref@jnl#1{{\jnl@style#1}}

\def\aaref@jnl#1{{\jnl@style#1}}

\def\aj{\aaref@jnl{AJ}}                   
\def\apj{\aaref@jnl{ApJ}}                 
\def\apjl{\aaref@jnl{ApJ}}                
\def\apjs{\aaref@jnl{ApJS}}               
\def\apss{\aaref@jnl{Ap\&SS}}             
\def\aap{\aaref@jnl{A\&A}}                
\def\aapr{\aaref@jnl{A\&A~Rev.}}          
\def\aaps{\aaref@jnl{A\&AS}}              
\def\mnras{\aaref@jnl{Mon.~Not.~Roy.~Astron.~Soc.}}             
\def\prd{\aaref@jnl{Phys.~Rev.~D}}        
\def\prc{\aaref@jnl{Phys.~Rev.~C}}  
\def\prl{\aaref@jnl{Phys.~Rev.~Lett.}}    
\def\qjras{\aaref@jnl{QJRAS}}             
\def\skytel{\aaref@jnl{S\&T}}             
\def\ssr{\aaref@jnl{Space~Sci.~Rev.}}     
\def\zap{\aaref@jnl{ZAp}}                 
\def\nat{\aaref@jnl{Nature}}              
\def\aplett{\aaref@jnl{Astrophys.~Lett.}} 
\def\apspr{\aaref@jnl{Astrophys.~Space~Phys.~Res.}} 
\def\physrep{\aaref@jnl{Phys.~Rep.}}      
\def\physscr{\aaref@jnl{Phys.~Scr}}       
\def\commat{\aaref@jnl{Comm.~Math.~Phys.}}              
\def\science{\aaref@jnl{Science}}               
\def\cqg{\aaref@jnl{Classical Quant.~Grav.}}            
\def\jpcs{\aaref@jnl{JPCS}}                                     
\def\ijmpd{\aaref@jnl{Int.~J.~Mod.~Phys.~D}}                    
\def\grg{\aaref@jnl{Gen.~Relat.~Gravit.}}               
\def\rpp{\aaref@jnl{Rep.~Prog.~Phys.}}          
\def\npa{\aaref@jnl{Nucl.~Phys.~A}}        
\def\lrr{\aaref@jnl{Living Rev.~Rel.}}                   
\def\jcap{\aaref@jnl{J.~Cosmology Astropart.~Phys.}}    
\def\rmp{\aaref@jnl{Rev.~Mod.~Phys.}}   
\def\epjc{\aaref@jnl{Eur.~Phys.~J.~C}} 
\def\plb{\aaref@jnl{~Phy.~Lett.~B}} 
\def\mpla{\aaref@jnl{Mod.~Phy.~Lett.~A}} 
\def\arxiv{\aaref@jnl{arxiv.org}}

\allowdisplaybreaks[1]

\begin{document}
\color{black}       
\title{\bf Rip cosmological models in extended symmetric teleparallel gravity}

\author{Laxmipriya Pati}
\email{lpriyapati1995@gmail.com}
\affiliation{Department of Mathematics,
Birla Institute of Technology and Science-Pilani, Hyderabad Campus,
Hyderabad-500078, India.}

\author{S.A. Kadam}
\email{k.siddheshwar47@gmail.com}
\affiliation{Department of Mathematics,
Birla Institute of Technology and Science-Pilani, Hyderabad Campus,
Hyderabad-500078, India.}

\author{S.K. Tripathy}
\email{tripathy\_sunil@rediffmail.com}
\affiliation{Department of Physics, Indira Gandhi Institute of Technology, Sarang, Dhenkanal, Odisha-759146, India.}

\author{B. Mishra}
\email{bivu@hyderabad.bits-pilani.ac.in}
\affiliation{Department of Mathematics,
Birla Institute of Technology and Science-Pilani, Hyderabad Campus,
Hyderabad-500078, India.}
\begin{abstract}
In this paper we have investigated some rip cosmological models in an extended symmetric teleparallel gravity theory. We consider the form $f(Q,T)=aQ^m+bT$ in the Einstein-Hilbert action and expressed the field equations and the dynamical parameters in terms of the non-metricity $Q$.  Three rip models such as Little Rip, Big Rip and Pseudo Rip are presented. The energy conditions and the cosmographic parameters are derived and analysed for all these models. 
\end{abstract}

\maketitle

\section{Introduction} 
The cosmological observations confirmed the presence of dark energy in the Universe and the change in its share in the mass energy budget of the Universe. Dark energy is believed to be responsible for the accelerated expansion of the Universe \cite{Riess98,Perlmutter99,Ade16,Aghanim20}. The concern arises among the cosmologists is that, by the expansion of the Universe, at a certain time in future the matter of the Universe will progressively torn apart. Cosmologists termed this situation as Big Rip (BR), which is a hypothetical model concerning the ultimate fate of the Universe. At finite cosmic time, the phantom energy density as well the scale factor become infinite \cite{Caldwell03,Frampton03,Nesseris04,Scherrer05}. To note here, the current observational data also supports the phantom dark energy model as well, in particular as the phantom dark energy, that may some way address the Hubble tension issue\cite{Valentino16,Vagnozzi20,Valentino21}.  In this scenario, the equation of state (EoS) parameter $\omega$ can be less than $-1$. In phantom energy dominated Universe, the dark energy operates as a phantom fluid, where we have $\omega<-1$, so in near future the occurrence of BR singularity can not be ruled out \cite{Sahni03}. In such a scenario, the evolution of the Universe will enter into a phantom phase and the expansion becomes super accelerating, which may lead to future singularity \cite{Nojiri05}. In standard matter, the decrease in scale factor resulted in the growth of energy density, however in the phantom energy dominated case, the energy density increases with the growth of the scale factor. Also the expansion of the Universe is based on the growth of the energy density. The BR singularity occurs even if all the energy conditions are violated. One important feature of the phantom model in comparison with the standard model is the phantom duality \cite{Dabrowski03}, where the field equations permit to map a small scale factor into a larger one and vice versa. Astashenok et al. \cite{Astashenok12} have shown the non-occurrence of finite time future singularity in the phantom dark energy model. So, the phantom dark energy models can be classified as: (i) cosmological constant, where the energy density evolves to a constant value in finite time; and (ii) phantom energy without BR singularity. Several aspects of BR singularity are available in the literature \cite{Starobinsky00}. \\

It appears that for the occurrence of singularity, the condition of EoS parameter less than $-1$ is not sufficient. The EoS parameter may asymptotically approach to $-1$ and the energy density increases or remains constant with the increase in time. So the occurrence of finite time future singularity may not occur \cite{Sahni03}. When the Hubble rate tends to constant, it may correspond to Pseudo Rip (PR) \cite{Astashenok12,Frampton12}. So, in PR, the density of dark energy increases monotonically with the scale factor, however bounded from above by some limiting density. For a sufficiently strong inertial force, this kind of models may lead to the dissolution of bound structure \cite{Frampton12}. Another type of rip cosmology is the Little Rip (LR) model, in which the energy density of the dark energy component increases monotonically with an unbounded from above and may not lead to a future singularity. However, this kind of models, ultimately lead to the dissolution of all bound structure \cite{Frampton11}. The dissolution of bound structure may not occur in BR and mostly depend on the model parameters, however it is inevitable in PR and LR models. The BR singularity that can occur not only in General Relativity (GR) but also in modified theories of gravity and seems to be the most terrifying future dark energy singularity.  The EoS parameter value from observational sources, Supernova data $\omega=-1.084 \pm 0.063$, WMAP $\omega=-1.073 \pm ^{0.090}_{0.089}$  favours $\Lambda$CDM model. The observational value of EoS parameter also favours the quintessence and phantom dark energy models. So, to avoid the future singularity, LR dark energy models are more appropriate, as the EoS parameter approaches $-1$ asymptotically and always finite at finite time.\\

Several cosmological models on rip cosmology are presented in modified theories of gravity. Gomez \cite{Gomez13} have shown the possibility of occurrence of LR and PR singularity due to the strength of the expansion. In the modified $F(R,G)$ gravity, in a reconstruction method, the effective phantom type model does not lead to future singularity \cite{Makarenko13}. Houndjo et al. \cite{Houndjo14} have shown that the second law of thermodynamics always satisfied around the LR Universe and obtained a stable fixed point. Also, the stability of the model has been assured with the linear perturbation analysis. The cosmological model with LR scale factor in $f(R,T)$ gravity resulted in the evolution of phantom like cosmic phase and overlap with the late time $\Lambda$CDM model \cite{Mishra20}. In the cosmological model, the LR abrupt is a cosmic doomsday, which can be prevented with the quantum effects and the classical evolution of the Universe in $f(R)$ gravity may avoid this doomsday with the DeWitt criterion \cite{Vasilev19,Vasilev21}. One can also experience a peculiar model with the big bang behaviour at the first half age of evolution, reaches BR and then with changing behaviour at the end of second half age of the Universe attained the big crunch \cite{Bakry21}. Another important result is that, with the present observational value of the deceleration parameter $q=-1.08$, the present value of the Hubble parameter $H_0$ for the BR and PR model can respectively be $74.33 Kms^{-1}Mpc^{-1}$ and $74.31 Kms^{-1}Mpc^{-1}$ \cite{Ray21}. In $f(T)$ gravity, obtaining the sequence of radiation, matter and late-time acceleration epochs, Hanafy and Saridakis \cite{Hanafy20} have observed that the Universe will be resulting in an everlasting PR phase. \\

The accelerated expansion of the Universe is responsible for the imbalance in the governing gravitational equations and hence GR is unable to address this issue. Several geometrical modified theories have come up in recent days, one among them is the $f(Q,T)$ gravity \cite{Xu19}, where $Q$ be the non-metricity and $T$ be the trace of energy momentum tensor. The $f(Q,T)$ gravity is originated from the teleparallel extension of GR. In the teleparallel representation, the curvature and non-metricity vanishes and the metric tensor $g_{\mu\nu}$ replaced by the set of tetrad vectors $e^i_{\mu}$. Nester and Yo \cite{Nester99} have suggested another equivalent representation, known as symmetric teleparallel gravity, where the geometric variable represented by $Q$ and ultimately framed as $f(Q)$ gravity \cite{Jimenez18}. Further, with the non-minimal coupling between the non-metricity and trace of energy momentum tensor the $f(Q,T)$ gravity has been framed \cite{Xu19}. The action for $f(T)$ and $f(Q)$ gravity are respectively $\int d^4x\sqrt{-g}T$  and $\int d^4x\sqrt{-g}Q$ and Jimenez et al. \cite{Jimenez19} have shown that in flat space both the actions are equivalent to that of GR.  We discuss here some of the recent astrophysical and cosmological results of $f(Q,T)$ gravity. The late time accelerating model in hybrid scale factor \cite{Pati21} and bouncing scenario \cite{Agrawal21} are discussed in $f(Q,T)$ gravity. The transient behaviour of the model with the general form of $f(Q,T)$ gravity can be obtained \cite{Zia21} and the model parameters for the linear case can be constrained with 31 Hubble data points and 57 Supernovae data \cite{Godani21}. The $f(Q,T)$ model can be reduced to $\Lambda$CDM model as a sub model with some specific values of the model parameters \cite{Najera21}.\\

The success of modified gravity in predicting the occurrence of rip singularity and the success of $f(Q,T)$ gravity in addressing the late time cosmic acceleration issue motivated us to investigate the occurrence of future singularity in $f(Q,T)$ gravity. The paper is organised as follows: in Sec. II, the action and basic field equations of $f(Q,T)$ gravity are given along with its dynamical parameters. In Sec. III, we have presented the energy conditions and the cosmographic parameters in terms of non-metricity. In Sec. IV, we have presented and analysed the Little, Big and Pseudo singularity models in $f(Q,T)$ gravity. Finally, the results and conclusion are given in Sec. V. 

\section{Field Equations and the Dynamical Parameters}

The systematic development from teleparallel gravity to symmetric teleparallel gravity, the $f(Q,T)$ gravity has been elaborated in the previous section. Further to develop the field equations of $f(Q,T)$ gravity the action needs to be defined, which can be given as \cite{Xu19},
\begin{equation} \label{eq.1}
S=\int\left[\dfrac{1}{16\pi}f(Q,T)d^{4}x\sqrt{-g}+\mathcal{L}_{m}d^{4}x\sqrt{-g}\right],
\end{equation}
Here, $f(Q,T)$ be the function of the non-metricity, $Q \equiv -g^{\mu \nu}( L^k_{~l\mu}L^l_{~\nu k}-L^k_{~lk}L^l_{~\mu \nu})$, where $L^k_{~l\gamma}\equiv-\frac{1}{2}g^{k\lambda}(\bigtriangledown_{\gamma}g_{l\lambda}+\bigtriangledown_{l}g_{\lambda \gamma}-\bigtriangledown_{\lambda}g_{l\gamma})$  and energy momentum tensor, $T=g^{\mu \nu}T_{\mu \nu}$. The determinant of the metric tensor and the matter Lagrangian respectively denoted as $g$ and $\mathcal{L}_m$ in the action. Now, varying the gravitational action \eqref{eq.1} with respect to the metric tensor, the field equation of $f(Q,T)$ gravity \cite{Xu19} can be obtained as, 
\begin{equation}\label{eq.2}
-\frac{2}{\sqrt{-g}}\bigtriangledown_{k}(f_{Q}\sqrt{-g}p^{k}_ {\mu \nu})-\frac{1}{2}fg_{\mu \nu}-f_{Q}(p_{\mu kl} Q^{\;\;\; kl}_{\nu}-2Q^{kl}_{\;\;\;\mu} p_{kl\nu})+f_{T}(T_{\mu \nu}+\Theta_{\mu \nu})=8 \pi T_{\mu \nu},
\end{equation}

In eqn. \eqref{eq.2}, we represent, $f\equiv f(Q,T)$ and $f_Q=\frac{\partial f}{\partial Q}$. The super potential of the model is defined as, $$p^{k}_{\mu \nu}=-\frac{1}{2}L^{k}_{\mu \nu}+\frac{1}{4}(Q^{k}-\tilde{Q}^{k})g_{\mu \mu}-\frac{1}{4}\delta^{k}_{(\mu}Q_{\nu)}.$$

We have the energy momentum tensor, $T_{\mu \nu}=\frac{-2}{\sqrt{-g}} \frac{\delta(\sqrt{-g}L_{m})}{\delta g^{\mu \nu}}$ and $\Theta_{\mu \nu}=g^{kl}\frac{\delta T_{kl}}{\delta g^{\mu \nu}}$  and the trace of the non-metricity as, $$Q_{k}=Q_{k}^{\;\;\mu}\;_{\mu}, ~~~~\tilde{Q}_{k}=Q^{\mu}\;_{k\mu}.$$

To construct the theoretical cosmological model of the Universe, we consider an isotropic and homogeneous FLRW space-time as,

\begin{eqnarray}\label{eq.3}
ds^{2}=-N^{2}(t)dt^{2}+\mathcal{R}^{2}(t)(dx^{2}+dy^{2}+dz^{2}),
\end{eqnarray}

In eqn. \eqref{eq.3}, the lapse function $N(t)$ considered to be 1 in the standard FLRW metric and $\mathcal{R}(t)$ be the scale factor which can be related to the Hubble function as, $H(t)=\frac{\dot{\mathcal{R}}(t)}{\mathcal{R}(t)}$. The dilation rate, $\tilde{T}=\frac{\dot{N}(t)}{N(t)}=0$, an over dot represents the derivative with respect to cosmic time and the non-metricity, $Q= 6H^2$. We consider here the energy momentum tensor in the form of perfect fluid as,
\begin{equation} \label{eq.4}
T_{\mu \nu}=diag(-\rho,p,p,p),  
\end{equation}
where $\rho$ and $p$ are respectively be the energy density and matter pressure and subsequently,
\begin{equation} \label{eq.5}
\Theta_{\mu \nu}=diag(2\rho+p,-p,-p,-p).
\end{equation}
We can obtain the $f(Q,T)$ gravity field equations \eqref{eq.2} for the FLRW space-time as,
\begin{eqnarray} 
f-12FH^2-4\dot{\chi}=-16\pi p, \label{eq.6} \\
f-12F H^2-4\dot{\chi}\kappa_1= 16\pi \rho, \label{eq.7}
\end{eqnarray}
where $F\equiv \frac{\partial f}{\partial Q}$,  $\chi=FH$, $\kappa_1=\frac{\kappa}{1+\kappa}$. From eqns. \eqref{eq.6} and \eqref{eq.7}, the evolution equation for the Hubble function can also be obtained as,

\begin{equation} \label{eq.8}
\dot{\chi}-4\pi\left(\rho+p\right)\left(1+\kappa\right)=0.
\end{equation}

Now, the equivalent Friedmann equations for the present gravity theory may be written as, 
\begin{eqnarray}
2\dot{H}+3H^2&=&\frac{1}{F}\left[\frac{f}{4}-2\dot{F}H+4\pi\left[(1+\kappa)\rho+(2+\kappa)p\right]\right]=-8\pi p_{eff},\label{eq.6a} \\
3H^2&=& \frac{1}{F}\left[\frac{f}{4}-4\pi\left[(1+\kappa)\rho+\kappa p\right]\right]=8\pi \rho_{eff}.  \label{eq.6b}
\end{eqnarray}
Obviously, the effective energy density $\rho_{eff}$ and the effective pressure $p_{eff}$ satisfy the conservation equation
\begin{equation} \label{eq.6c}
\dot{\rho}_{eff}+3H\left(\rho_{eff}+p_{eff}\right)=0.
\end{equation}

To understand the different stages of the Universe as well to study the claim of accelerated expansion of the Universe, the equation of state (EoS) parameter  $\omega$, which is the ratio of pressure and energy density plays a major role both theoretically and observationally. Therefore here we express the EoS parameter in terms of Hubble function as,

\begin{equation} \label{eq.9}
\omega=-1+\frac{4\dot{\chi}}{\left(1+\kappa\right)\left(f-12F H^2\right)-4\dot{\chi}\kappa}.
\end{equation}

Xu et al. \cite{Xu19} have proposed three functional form for the function $f(Q,T)$ as: (i) $f(Q,T)=aQ+bT$, (ii) $f(Q,T)=aQ^{m}+bT$ and (iii) $f(Q,T)=-\left(aQ+bT^2\right)$, where $a$, $b$ and $m$ are constants. We can see that, for $m=1$, $f(Q,T)=aQ^m+bT$ can reduce $f(Q,T)=aQ+bT$. We will proceed with the model by assuming the form, $f(Q,T)=aQ^{m}+bT$, which describes the non-minimal coupling between the non-metricity and trace of energy momentum tensor. With this eqns. \eqref{eq.6}, \eqref{eq.7} and \eqref{eq.9} can be respectively obtained as,

\begin{eqnarray}
p &=& \frac{-\left(1-2m\right)aQ^m+2\dot{\chi}\left[2+\kappa-\kappa\kappa_1\right]}{16\pi(1+2 \kappa)},  \label{eq.10}\\
\rho &=& \frac{(1-2m)aQ^m+2\dot{\chi}\left[3\kappa-\left(2+3\kappa\right)\kappa_1\right]}{16\pi(1+2\kappa)},  \label{eq.11}\\
\omega&=&\frac{-\left(1-2m\right)aQ^m+2\dot{\chi}\left[2+\kappa-\kappa\kappa_1\right]}{(1-2m)aQ^m+2\dot{\chi}\left[3\kappa-\left(2+3\kappa\right)\kappa_1\right]}.  \label{eq.12}
\end{eqnarray}

One of the major drawback in GR framework to address the cosmic acceleration issue is the violation of strong energy condition. This is inevitable since the EoS parameter needs to be negative during the evolution phase. Another important aspect of the cosmological model is on its geometrical validation. So, in the next section, we shall present the general form of energy conditions and different geometrical parameters.
\section{Energy Conditions and Cosmographic Parameters}
The energy conditions as defined in GR can also be defined in the context of $f(Q,T)$ gravity with the new pressure and energy density term obtained respectively in eqns. \eqref{eq.10} and \eqref{eq.11}. The energy conditions are based on Raychaudhuri equations and it can be described by the behaviour of a congruence of space-like, time-like or light-like curves \cite{Hawking99}. These conditions give better understanding on the singularity theorem, geodesics of the Universe and so on. When the matter is in the form of perfect fluid, the four energy conditions can be given as, null energy condition: $\rho+p\geqslant0$, weak energy condition: $\rho+p\geqslant0$, $\rho\geqslant0$, the strong energy condition: $\rho+3p\geqslant0$, and the dominant energy condition: $\rho-p\geqslant0$. Since the violation of strong energy condition has become essential in the context of modified theories of gravity, its survival is now at stake. Now, the energy conditions such as NEC, WEC, SEC and DEC for this $f(Q,T)$ gravity model can be expressed as,

\begin{eqnarray}
\rho+p&=&\frac{1}{4\pi}\left[(1-\kappa_1) \dot{\chi}\right],\nonumber \\ 
\rho+p&=&\frac{1}{4\pi}\left[(1-\kappa_1) \dot{\chi}\right], \quad \rho>0, \nonumber\\
\rho+3p&=&\frac{1}{8\pi}\left[\frac{-(1-2m)aQ^m+(6+6\kappa-2\kappa_1-6\kappa \kappa_1)\dot{\chi}}{(1+2\kappa)}\right],\nonumber\\
\rho-p&=&\frac{1}{8\pi}\left[\frac{(1-2m)aQ^m-2(1-\kappa+\kappa_1+\kappa \kappa_1)\dot{\chi}}{(1+2\kappa)}\right] \label{eq.13}.
\end{eqnarray}

Cosmography uses symmetries to derive the FLRW form of the metric. It can view the history of the scale factor $\mathcal{R}(t)$, which can be treated as a free quantity to be determined observationally. Then the thought process should go into the dynamics of the model. Practically, this involves in satisfying a finite number of derivatives of the scale factor. The Taylor series expansion handled this finite derivatives at present epoch value as,

\begin{equation}
\mathcal{R}=\mathcal{R}_0\left[1+H_0(t-t_0)-\frac{q_0}{2!}H_0^2(t-t_0)^2+\frac{j_0}{3!}H_0^3(t-t_0)^3+\frac{s_0}{4!}H_0^4(t-t_0)^4+\frac{l_0}{5!}H_0^5(t-t_0)^5+O([t-t_0]^6)\right],\label{eq.14}
\end{equation}
where $R_0$ be the scale factor at present epoch. $H_0$, $q_0$, $j_0$, $s_0$ and $l_0$ respectively denotes Hubble parameter, deceleration parameter, jerk parameter, snap parameter and lerk parameter at time $t_0$. These parameters enable us to study the past and future behaviour of the Universe. The deceleration parameter decides the accelerating behaviour of the model. The $(j,s)$ pair discriminate between dark energy models. The exact behaviour of the lerk parameter is yet to be ascertained. We shall express these geometrical parameters in terms of the scale factor as,
\begin{eqnarray}
H(t)&=&\frac{1}{\mathcal{R}}. \frac{d \mathcal{R}}{dt}, \nonumber\\
q(t)&=&-\frac{1}{\mathcal{R}}\frac{d^2\mathcal{R}}{dt^2} \left[\frac{1}{\mathcal{R}}\cdot \frac{d \mathcal{R}}{dt}, \right]^{-2} \nonumber\\
j(t)&=&\frac{1}{\mathcal{R}}\frac{d^3\mathcal{R}}{dt^3} \left[\frac{1}{\mathcal{R}}\cdot \frac{d \mathcal{R}}{dt} \right]^{-3}, \nonumber\\
s(t)&=&\frac{1}{\mathcal{R}}\frac{d^4\mathcal{R}}{dt^4} \left[\frac{1}{\mathcal{R}}\cdot \frac{d \mathcal{R}}{dt} \right]^{-4},\nonumber\\
l(t)&=&\frac{1}{\mathcal{R}}\frac{d^5\mathcal{R}}{dt^5} \left[\frac{1}{\mathcal{R}}\cdot \frac{d \mathcal{R}}{dt} \right]^{-5} \label{eq.15}
\end{eqnarray}

The energy conditions and geometrical behaviour would be analysed in the respective models in the following section.

\section{The Models}
In fact, the choices of the scale factors will satisfy the field equations. Here, in stead of choosing a specific expression of the EoS parameter in terms of $p=\omega\rho$, we have considered the Hubble parameter that satisfies the field equations. Usually, it is possible to solve exactly the field equations of the symmetric teleparallel gravity theory for a given choice of $\omega$, to obtain an expression of the scale factor. Here, we chose the other way, considered a scale factor and obtained an expression for the EoS parameter which obviously bears a dynamical nature. \\

The motivation is to investigate the future evolution of the Universe through LR, BR and PR scale factors, since the accelerated expansion of the Universe may lead to the possibility of breaking of all bound structures, including the space-time. We shall investigate here the cosmological models based on LR, BR and PR scale factor.

\subsection{LR Model}
\begin{figure} [H]
\centering
\includegraphics[width=85mm]{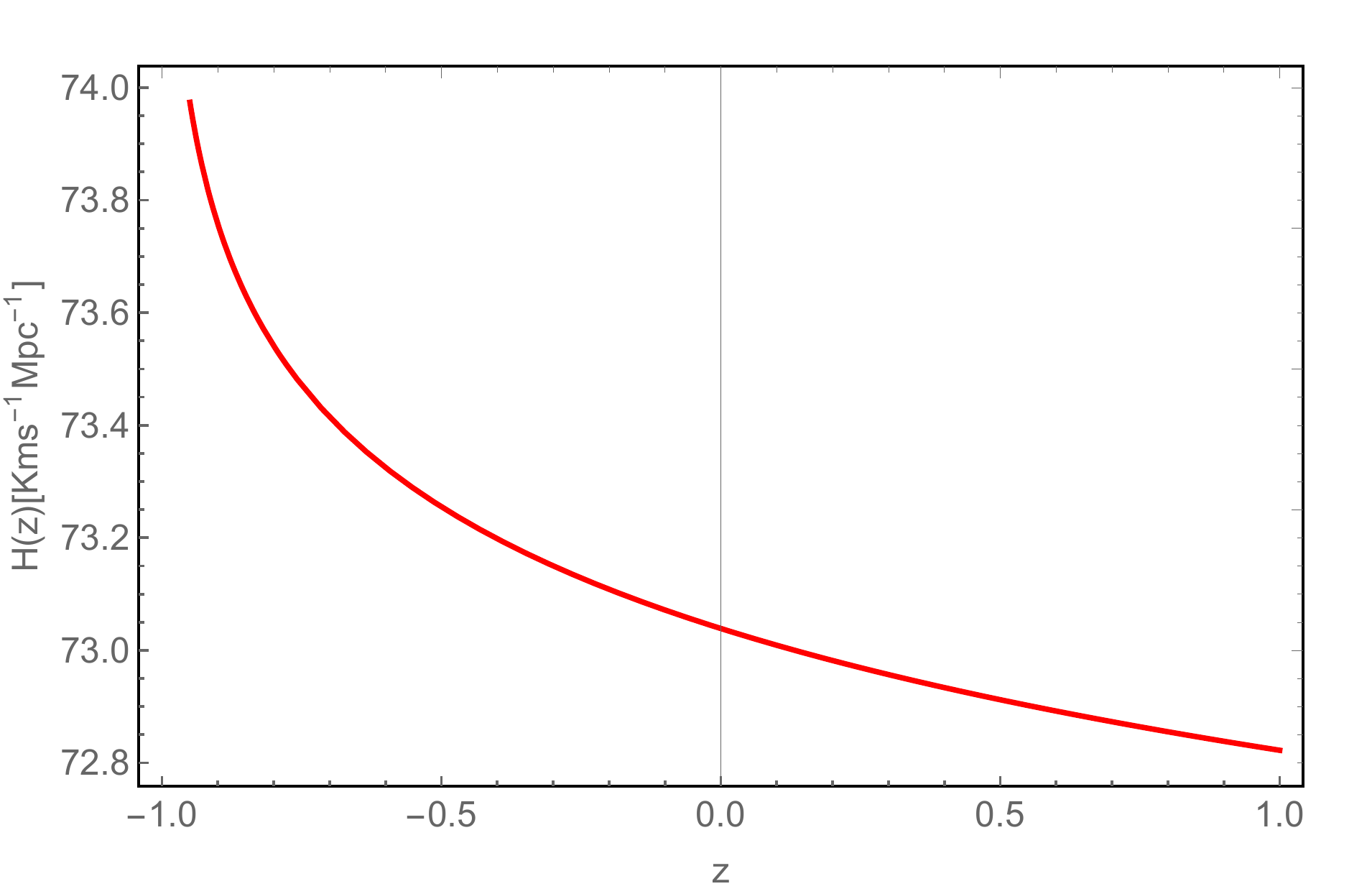}
\includegraphics[width=85mm]{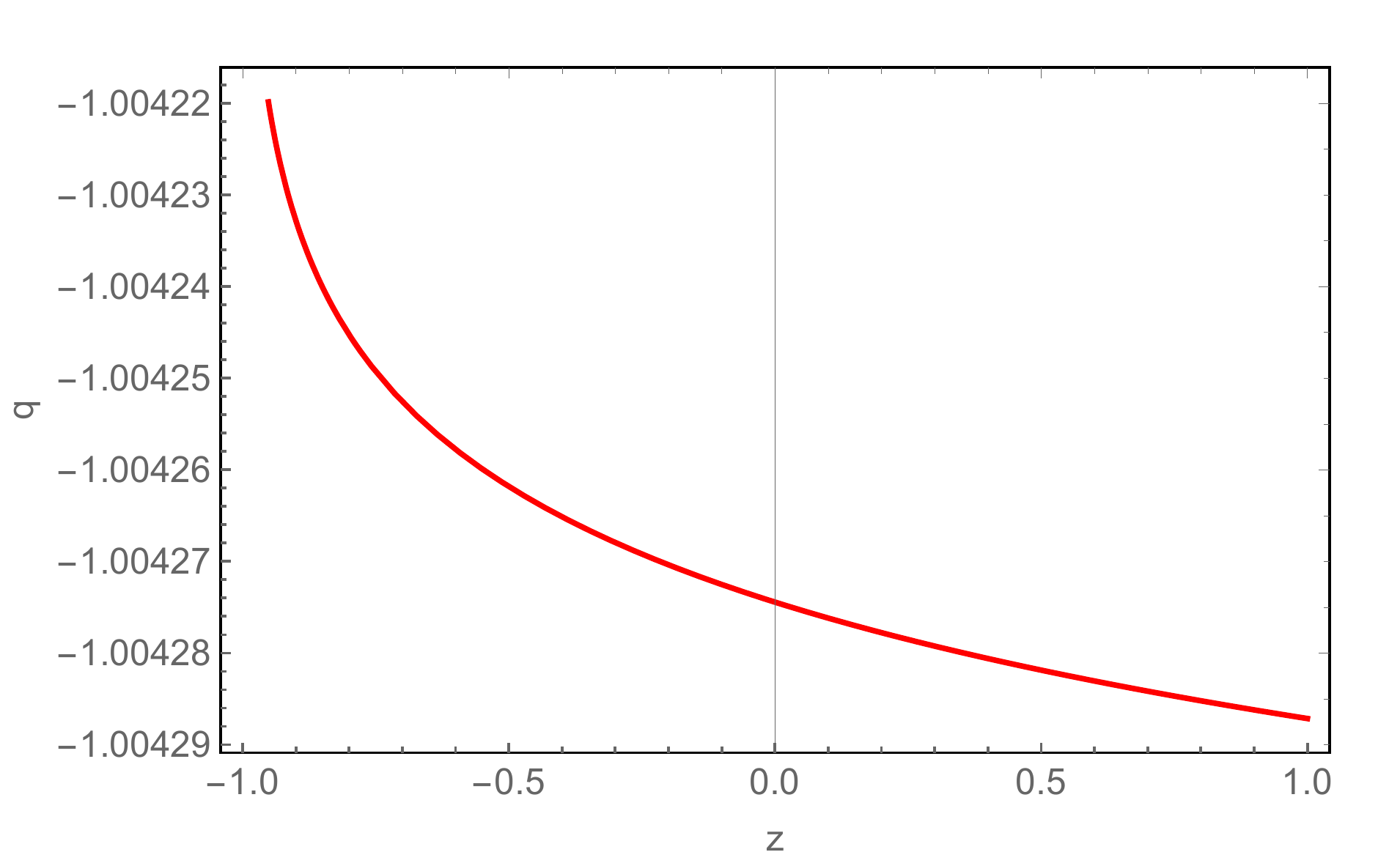}
\caption{Behaviour of Hubble parameter (left panel) and deceleration parameter (right panel) in redshift, ($A=25.11$, $\lambda=0.3122$, $t_0=3.42$).} \label{FIG1}
\end{figure}

\begin{figure} [H]
\centering
\includegraphics[width=85mm]{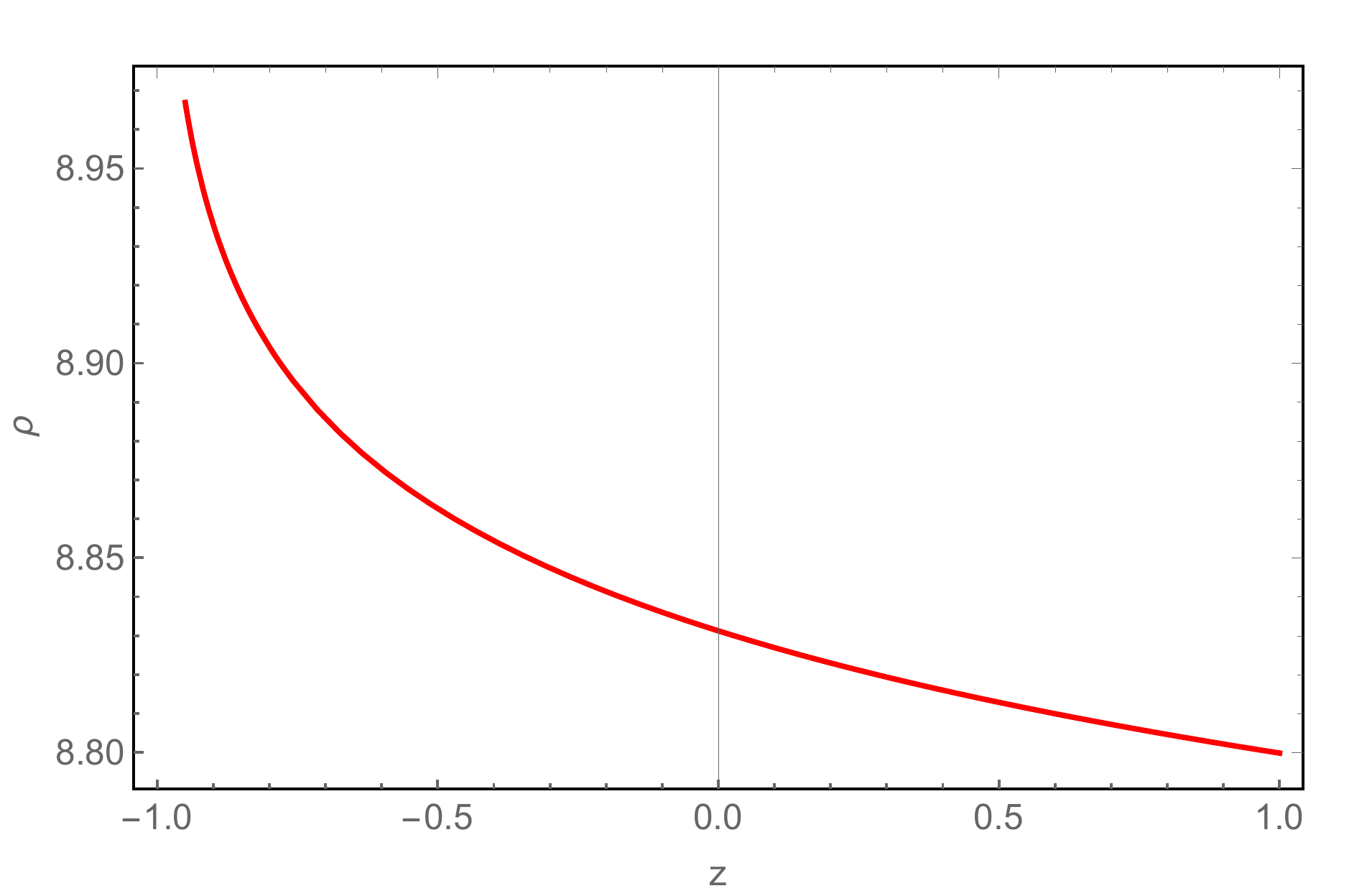}
\includegraphics[width=85mm]{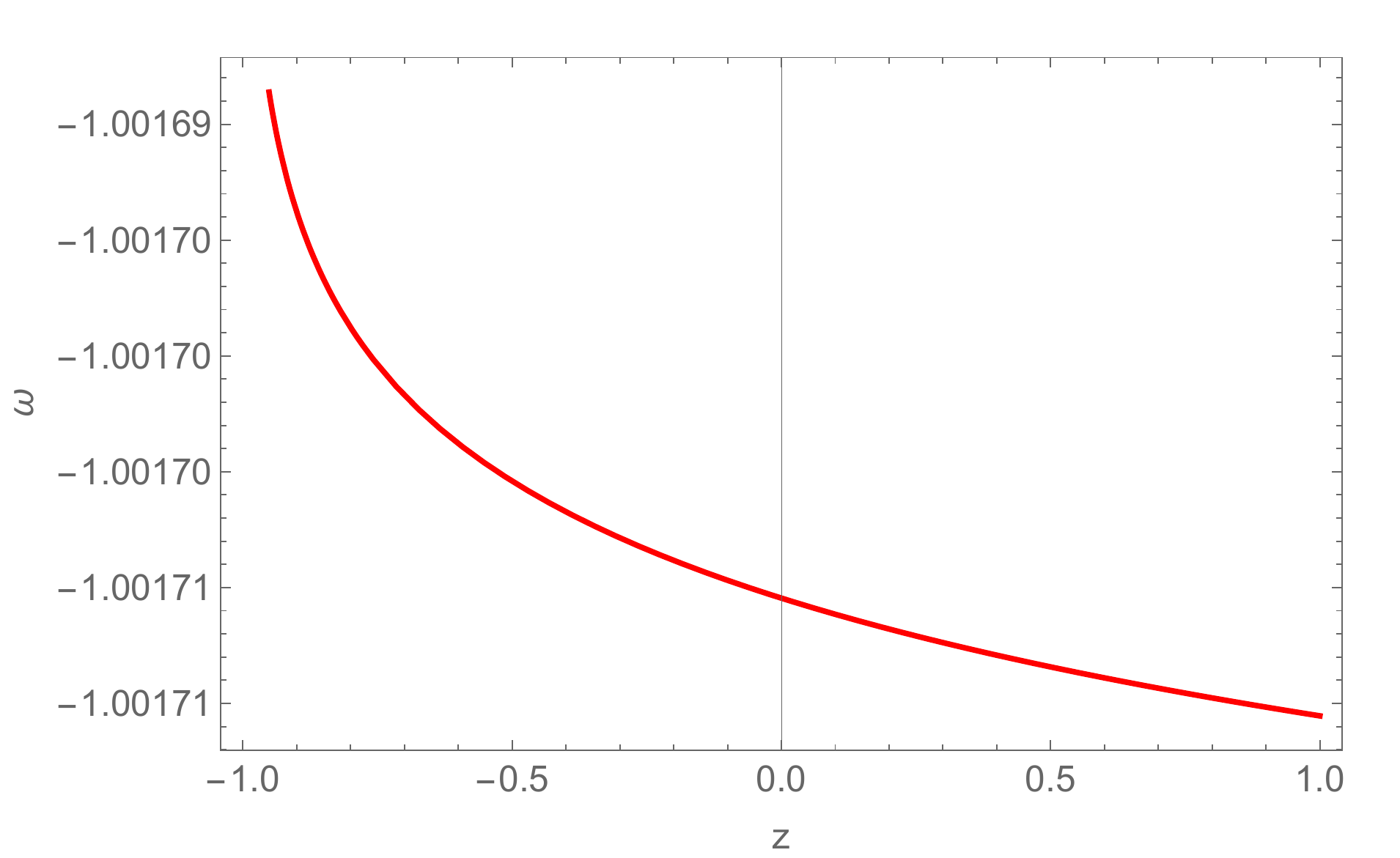}
\caption{Behaviour of energy density (left panel) and EoS parameter (right panel) in redshift, ($a=-4.4$, $b=0.01$, $m=0.6$, $A=25.11$, $\lambda=0.3122$, $t_0=3.42$).} \label{FIG2}
\end{figure}

We consider the LR scale factor in the form,
\begin{equation}
\mathcal{R}=\mathcal{R}_0 exp\left[\frac{A}{\lambda}(e^{\lambda t}-e^{\lambda t_0})\right], \label{eq.16}
 \end{equation}
where $A$ and $\lambda$ are positive constants. The Hubble parameter for this model becomes, 
$H = Ae^{\lambda t}$ and apparently, the LR behaviour can be obtained for $\lambda>0$. The scale factor at present epoch is considered as $\mathcal{R}_0=1$. The deceleration parameter becomes, $q=-1-\frac{\lambda e^{-\lambda t}}{A} $. The Hubble rate diverges when $t \rightarrow \infty$ and with time, it increases exponentially FIG. \ref{FIG1} (left panel). With suitable choice of a parameter space, the present value of the Hubble parameter is obtained as $\approx 73.03 Kms^{-1}Mpc^{-1}$. The deceleration parameter throughout the evolution remains negative, for the positive value of $\lambda$ and $A$. It increases from a higher negative value and remains $-1$ at late time of the evolution. The deceleration parameter always remains negative for positive value of $\lambda$ since $e^{-\lambda t}$ is positive. So, to experience a decelerating Universe, an appropriate negative value of $\lambda$ can be considered. So, the accelerating or deceleration Universe can be experienced with the sign of the scale factor parameter. FIG. \ref{FIG1} (right panel) depicts the behaviour of deceleration parameter for the accelerating era only and at present epoch $q_0=-1.016$ close to the value of the analysis $q_0=-1.08\pm 0.29$ \cite{Camarena20}. At finite time, the LR model avoids singularity, hence the transient behaviour of the deceleration parameter can not be demonstrated.  
The accelerating behaviour of the cosmological model can be assessed through the EoS parameter. So we present here the dynamical parameters of the LR model in $f(Q,T)$ gravity from eqns.\eqref{eq.10}-\eqref{eq.12} as,    
\begin{eqnarray}
p&=& -\frac{a(6)^{m-1}(1-2m)(A e^{\lambda t})^{2m-1}}{8\pi(1+2\kappa)}\left[3A e^{\lambda t}+m\lambda(2+\kappa-\kappa \kappa_1)\right], \label{eq.17}\\
\rho&=&\frac{a(6)^{m-1}(1-2m)(A e^{\lambda t})^{2m-1}}{8\pi(1+2\kappa)}\left[3A e^{\lambda t}-m\lambda(3\kappa-2\kappa_1-3\kappa \kappa_1)\right], \label{eq.18}\\
\omega&=&-1-\frac{2m\lambda(1-\kappa_1+2 \kappa-2\kappa \kappa_1)}{3A e^{\lambda t}-m\lambda(3\kappa-2\kappa_1-3\kappa \kappa_1)}. \label{eq.19}
\end{eqnarray}
The behaviour of these dynamical parameters behaviour depend on the parameters $a$, $b$, $m$, $\lambda$ and $A$. For an accelerating behaviour, $\lambda=0.3122 (Gyr)^{-1}$ is taken to keep the deceleration parameter in the preferred range. The other model parameters are chosen in such a manner that the energy density remains positive throughout [FIG. \ref{FIG2} (left panel)] and an accelerating behaviour of the EoS parameter can be achieved. Since the exponent $m$ has a role in the functional form of the prescribed $f(Q,T)$, we have chosen here the values of $m=0.6$. To mention for $m=1$, the behaviour would reduce to the linear case. The EoS parameter evolves dynamically in the phantom region $\omega<-1$, gradually increases to reach the concordance value, $\omega=-1$ at late times. So, at late times of the cosmic evolution, the LR model overlaps with the $\Lambda$CDM model [FIG. \ref{FIG2} (right panel)]. From eqn. \eqref{eq.19}, we can assess that the value of $m$ affects the evolution behaviour of the EoS parameter, i.e. the growth rate is low with the higher value of $m$. However, at late times, the second term would vanish and $\omega$ would be asymptotically merged to $-1$. For $m=0.6$, the present value of EoS parameter is obtained as $\omega_0=-1.001$, which is close to that of the concordant $\Lambda$CDM Universe.

\begin{figure} [H]
\centering
\includegraphics[width=85mm]{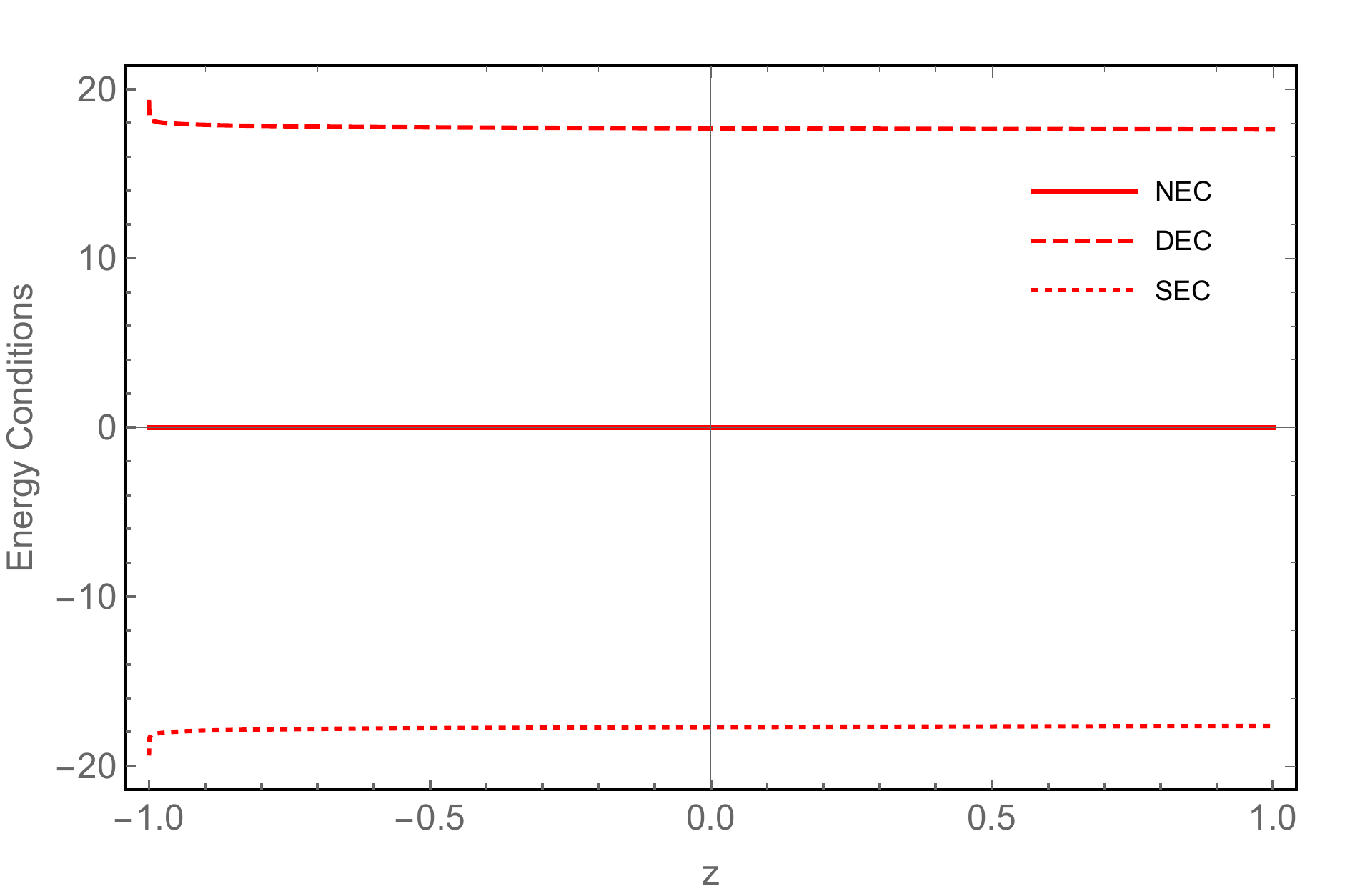}
\includegraphics[width=85mm]{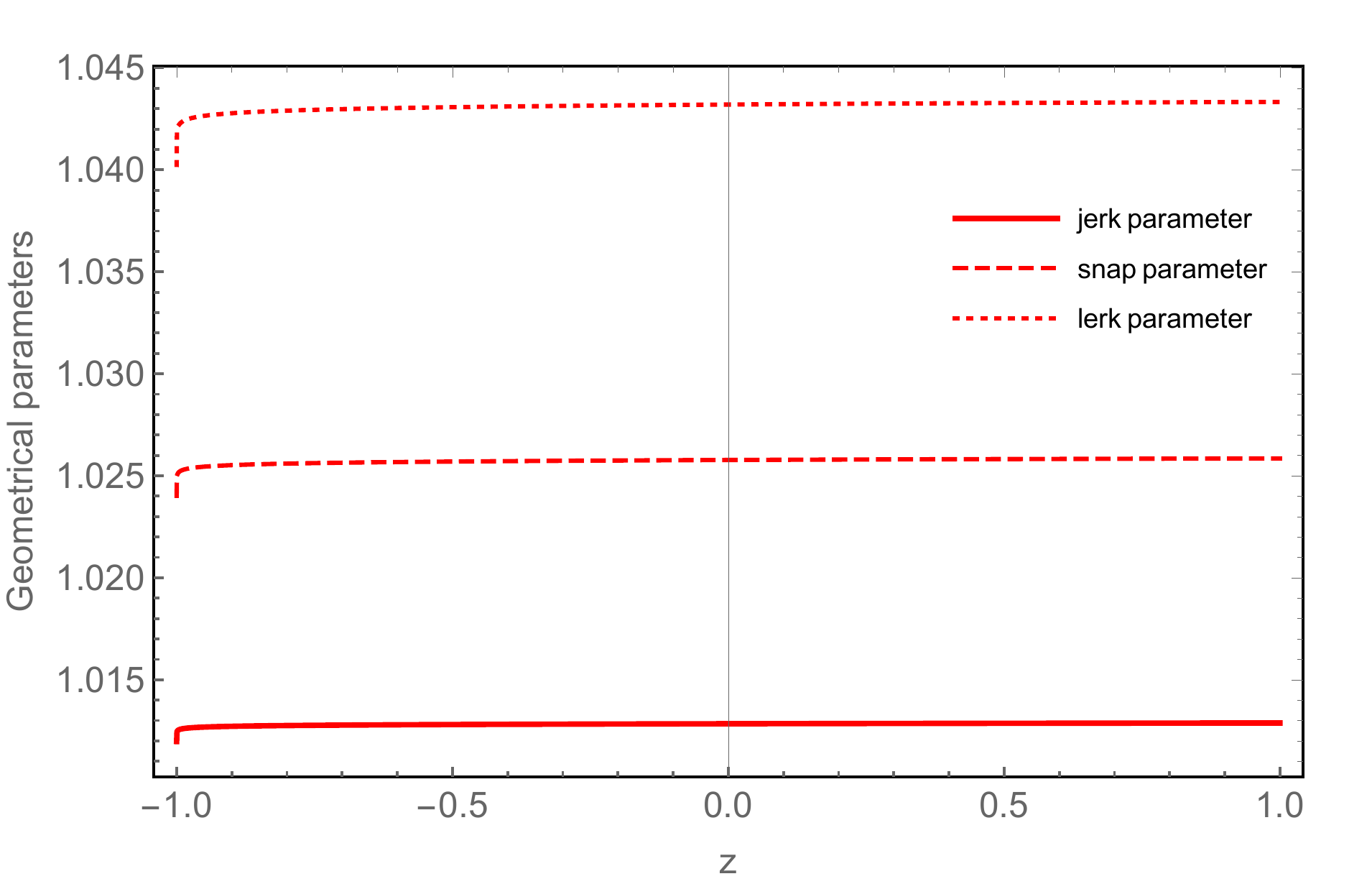}
\caption{Behaviour of energy conditions (left panel) and geometrical parameters (right panel) in redshift, ($a=-4.4$, $b=0.01$, $m=0.6$, $A=25.11$, $\lambda=0.3122$, $t_0=3.42$).} \label{FIG3}
\end{figure}

The energy conditions \eqref{eq.13}  for LR case can be reduced to,

\begin{eqnarray}
\rho+p &=& -\frac{a (6)^{m-1} m(1-2m) \lambda (Ae^{\lambda t})^{2m-1}} {4\pi}\left[1-\kappa_1\right], \nonumber \\
\rho+3p&=&-\frac{a (6)^{m-1} (1-2m) (Ae^{\lambda t})^{2m-1}} {4\pi(1+2\kappa)}\left[3Ae^{\lambda t}+m \lambda(3+3\kappa-\kappa_1-3\kappa \kappa_1)\right], \nonumber \\
\rho-p&=&\frac{a (6)^{m-1} (1-2m) (Ae^{\lambda t})^{2m-1}} {4\pi(1+2\kappa)}\left[3Ae^{\lambda t}+m \lambda(1-\kappa+\kappa_1+\kappa \kappa_1)\right].\label{eq.20}
\end{eqnarray}
The graphical behaviour of energy conditions are given in FIG. \ref{FIG3} (left panel). We can infer that with the change in the value of $m$ the behaviour of the EoS parameter remains same, but there might be some change in the stiffness. This would not affect the behaviour of the energy conditions. So, here we consider the same value of the exponent, $m=0.6$. The NEC remains constant throughout and mostly lying on the null line. The violation of SEC further validates the accelerating behaviour of the Universe whereas the DEC does not violate as expected. The SEC and DEC in their respective domain slightly move away at late times of the evolution.     
 
The cosmographic parameters for the LR model are,
\begin{eqnarray}
j&=&1+3\frac{\lambda e^{-\lambda t}}{A}+\frac{{\lambda}^{2}e^{-2\lambda t}}{A^{2}}, \nonumber\\
s&=& 1+6\frac{\lambda e^{-\lambda t}}{A}+7\frac{{\lambda}^{2}e^{-2\lambda t}}{A^{2}}+\frac{{\lambda}^{3}e^{-3\lambda t}}{A^{3}}, \nonumber\\
l&=&1+10\frac{\lambda e^{-\lambda t}}{A}+25\frac{{\lambda}^{2}e^{-2\lambda t}}{A^{2}}+15\frac{{\lambda}^{3}e^{-3\lambda t}}{A^{3}}+\frac{{\lambda}^{4}e^{-4\lambda t}}{A^{4}}. \label{eq.21}
\end{eqnarray}
The state finder pair $(j,s)$ approaches to $(1,1)$ at late times thereby indicate the SCDM behaviour. The lerk parameter reduces from higher value and approaches to unity. The jerk parameter evolves from a lower value as compared to the snap parameter and the same behaviour is preserved between snap and lerk parameter.

\subsection{The BR Model}

In the big rip, the scale factor and density diverge in a singularity at a finite future time. Due to the nature of constant dark energy density remains constant there is no such divergence and no disintegration.  A feature of a big rip is that all bound-state systems disintegrate before the final singularity. Here we consider a scale factor as, 
\begin{equation}
\mathcal{R}=R_0+\frac{1}{(t_s-t)^{\alpha}}, \label{eq.22}
\end{equation}
where $t_s$ and $\alpha$ are free parameters and can be constrained from the physical basis and $R_0$ be the integrating constant. The Hubble parameter, $H=\frac{\alpha}{(t_s-t)[\mathcal{R}_0(t_s-t)^{\alpha}+1]}$, and the BR situation can arise at $t=t_s$. To note, when the integration constant, $\mathcal{R}_0=0$, the Hubble parameter reduces to,  $H=\frac{\alpha}{t_s-t}$. It can be observed that, $\dot{H}$ remains positive irrespective of phantom-like or non-phantom-like phase. The deceleration parameter can be obtained as,  $q=-\frac{(\alpha+1)\left(1+R_{0}(t_{s}-t)^{\alpha}\right)}{\alpha}$.

\begin{figure} [H]
\centering
\includegraphics[width=85mm]{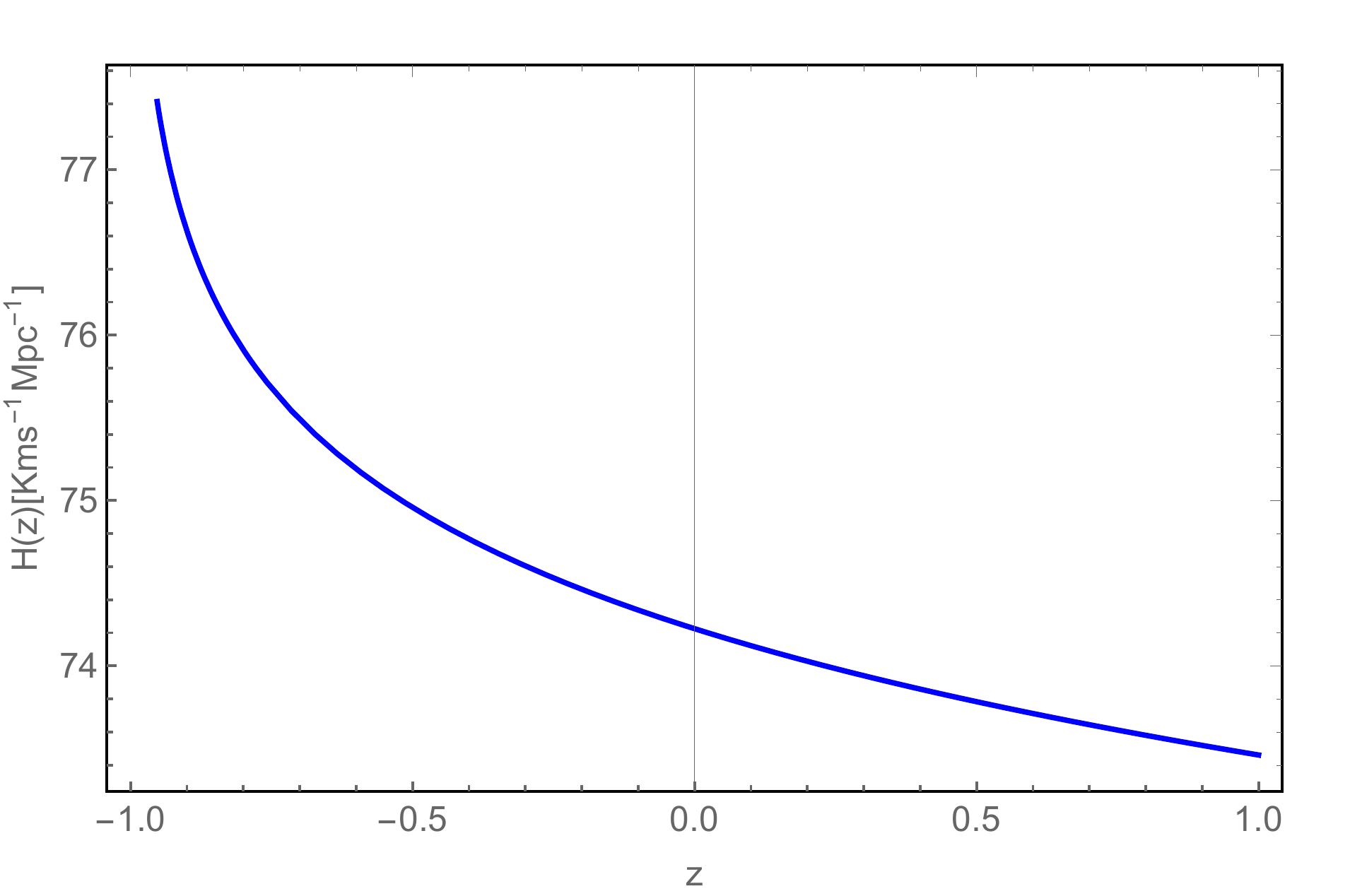}
\includegraphics[width=85mm]{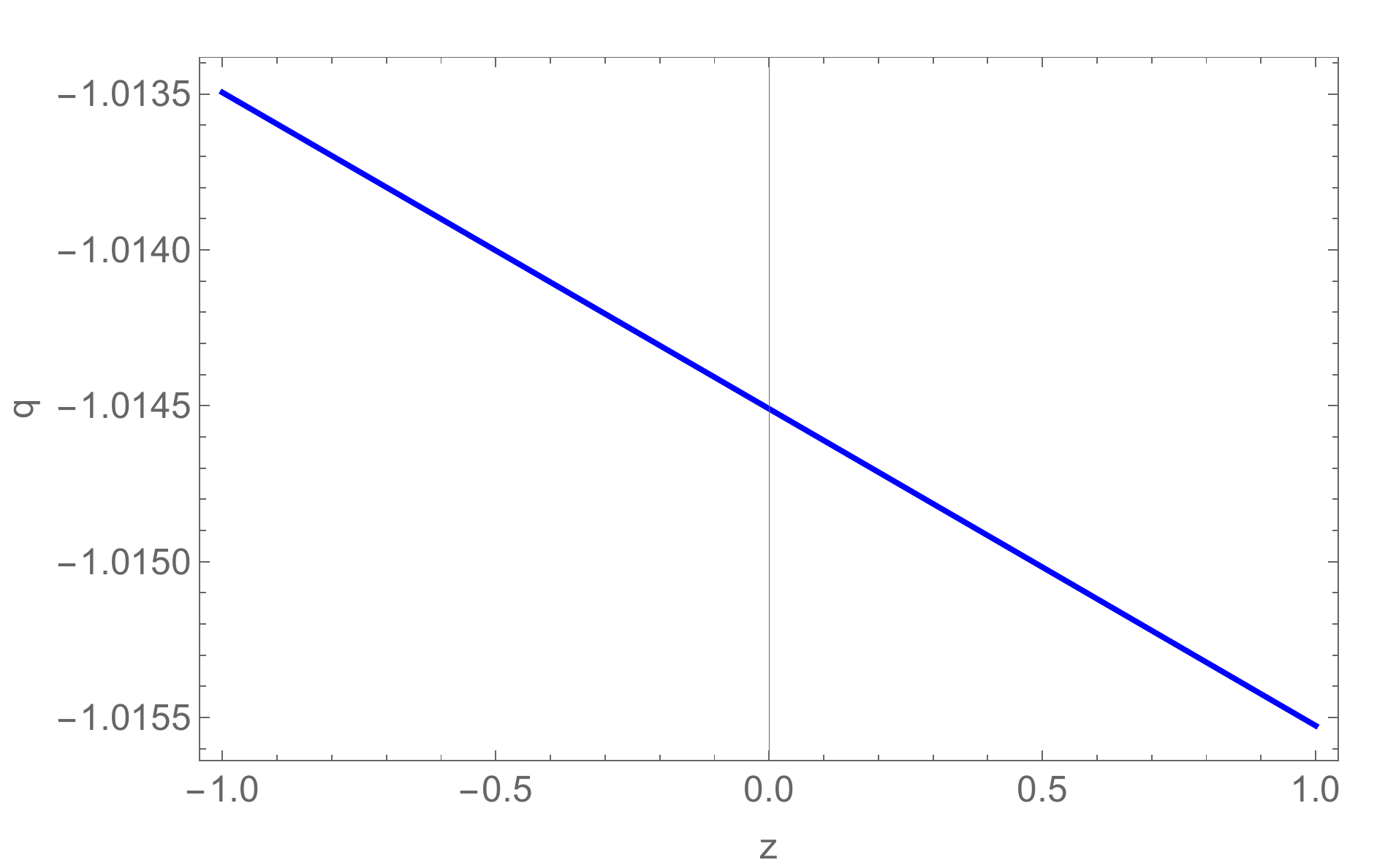}
\caption{Behaviour of Hubble parameter (left panel) and deceleration parameter (right panel) in redshift, ($\alpha=74.1$, $t_s=13.8$, $\mathcal{R}_0=0.001$).} \label{FIG4}
\end{figure}

The Hubble parameter increases over the time and at present time, $H_0=74.22 Kms^{-1}Mpc^{-1}$ [FIG. \ref{FIG4} (left panel)]. At the same time the deceleration parameter increases in the negative domain and at late time stay very close to $-1$. At the present time the parameter can be observed to be, $q_0\approx-1.015$ and thereby shows the accelerating behaviour [FIG. \ref{FIG4} (right panel)]. We have constrained the moment $t_s$ from some physical basis. Now, for the BR model the dynamical parameters can be obtained as,

\begin{eqnarray}
p&=& -\frac{a(6)^{m-1}(1-2m)\left(\frac{\alpha}{t_s-t}\right)^{2m-2}}{8\pi(1+2\kappa)}\left[3\left(\frac{\alpha}{t_s-t}\right)^2+m(2+\kappa-\kappa \kappa_1)\frac{\alpha}{(t_s-t)^2}\right], \label{eq.23}\\
\rho &=&\frac{a(6)^{m-1}(1-2m)\left(\frac{\alpha}{t_s-t}\right)^{2m-2}}{8\pi(1+2\kappa)}\left[3\left(\frac{\alpha}{t_s-t}\right)^2-m(3\kappa-2\kappa_1-3\kappa \kappa_1)\frac{\alpha}{(t_s-t)^2}\right], \label{eq.24}\\
\omega &=& -1-\frac{2m(1-\kappa_1+2\kappa-2\kappa \kappa_1)\frac{\alpha}{(t_s-t)^2}}{3\left(\frac{\alpha}{t_s-t}\right)^2-m(3\kappa-2\kappa_1-3\kappa \kappa_1)\frac{\alpha}{(t_s-t)^2}}.\label{eq.25}
\end{eqnarray}
\begin{figure} [H]
\centering
\includegraphics[width=85mm]{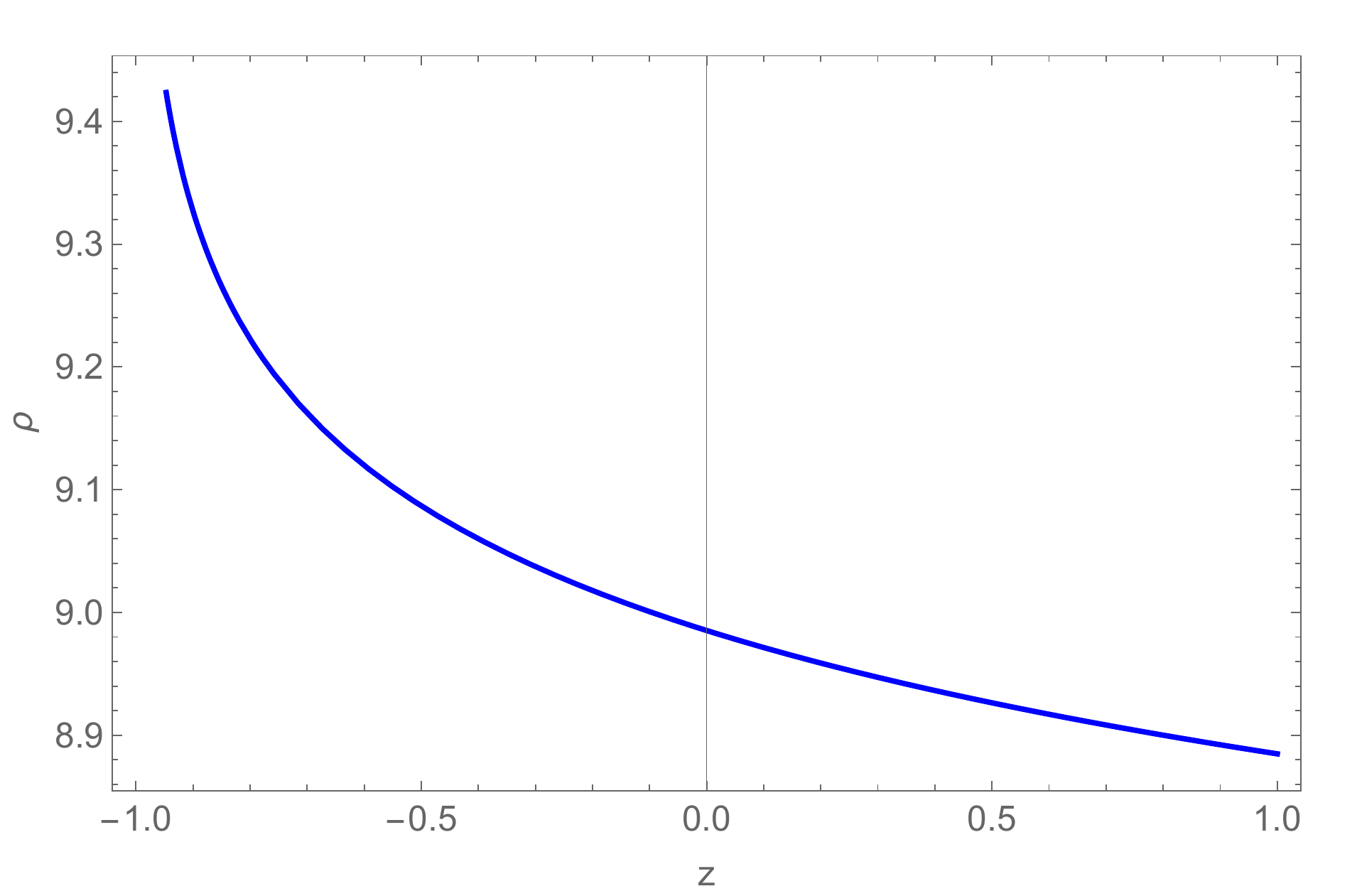}
\includegraphics[width=85mm]{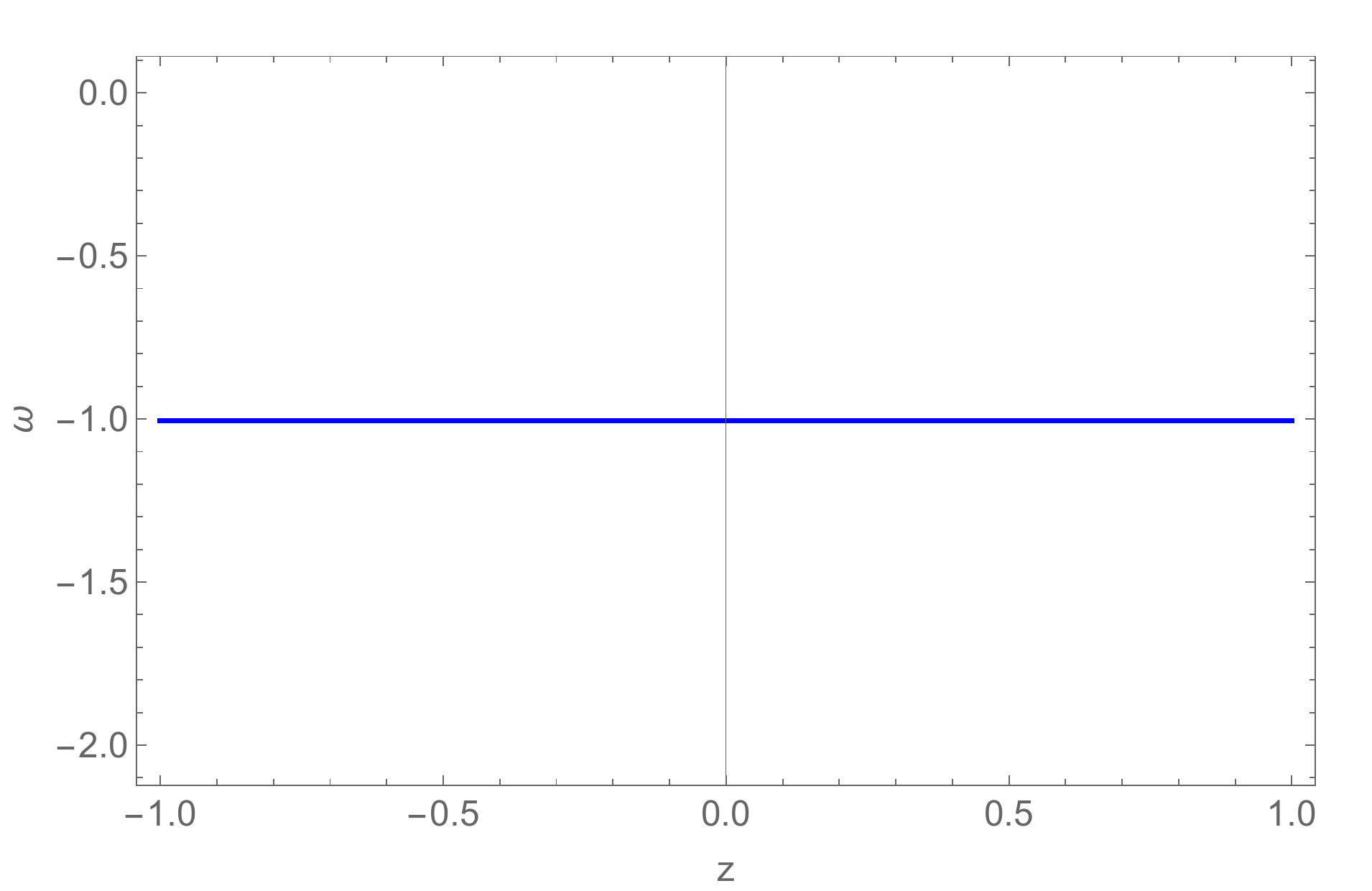}
\caption{Behaviour of energy density (left panel) and EoS parameter(right panel) in redshift, ($a=-4.4$, $b=0.01$, $m=0.6$, $\alpha=74.1$, $t_s=13.8$, $\mathcal{R}_0=0.001$).} \label{FIG5}
\end{figure}
The dynamical parameters of this model characterizes the accelerating behaviour of the Universe along with the possible occurrence of the BR singularity. With the physically constrained values of the parameters, the matter pressure remains throughout  negative. At the same time, during the entire evolution, the energy density [FIG. \ref{FIG5} (left panel)] observed to be positive and since there is no observational evidences on the present value of the energy density, we have to depend on the present value of the EoS parameter to validate the model. The EoS parameter throughout remain constant at $-1$. Since $\omega=-1$ shows the $\Lambda$CDM behaviour of the Universe, we claim this model as accelerating and shows $\Lambda$CDM behaviour [FIG. \ref{FIG5} (right panel)]. The occurrence of BR can be avoided in this model with this suitable choice of the parameters value. 

\begin{figure} [H]
\centering
\includegraphics[width=85mm]{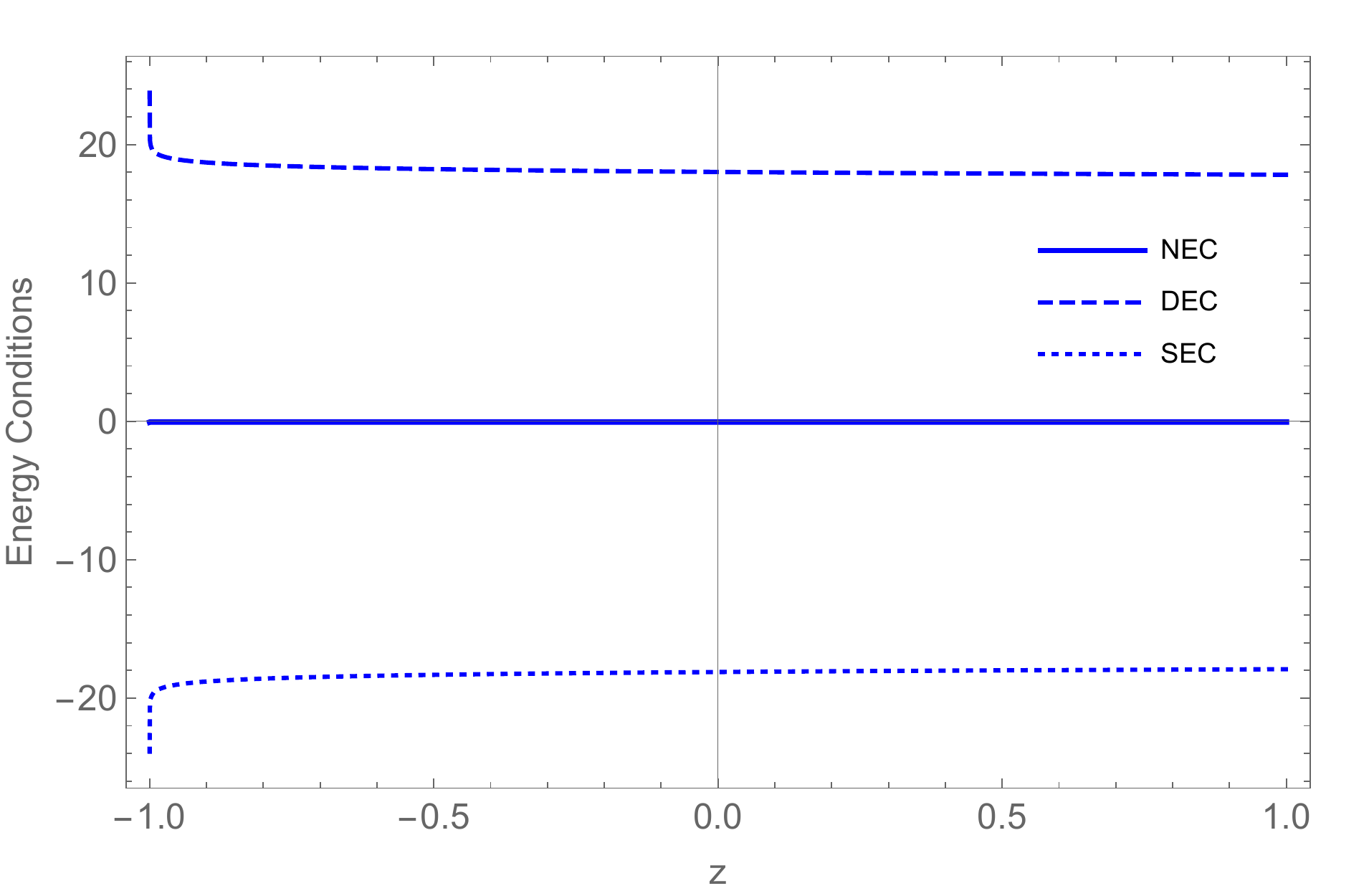}
\includegraphics[width=85mm]{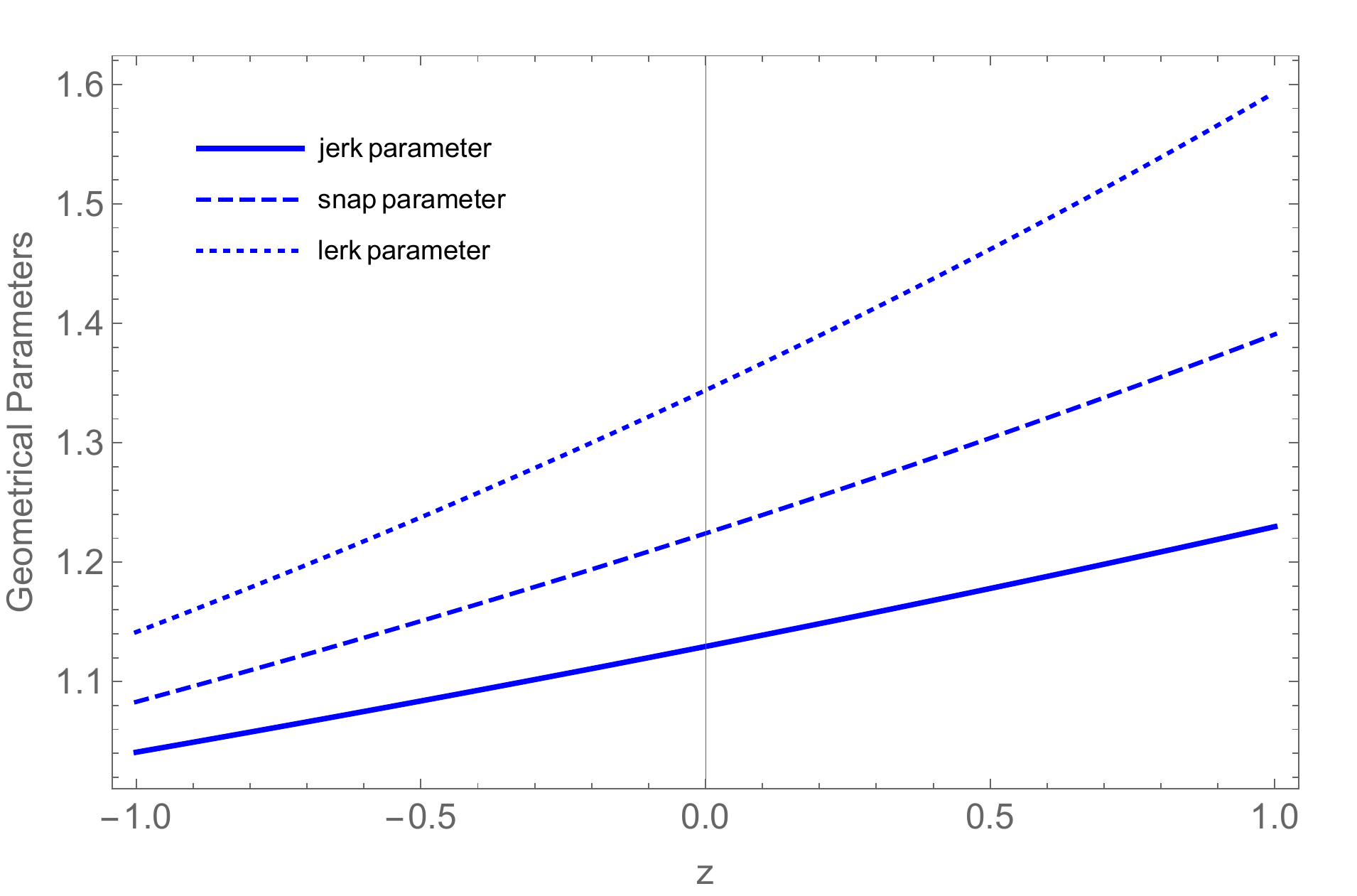}
\caption{Behaviour of energy conditions (left panel) and geometric  parameters(right panel) in redshift, ($a=-4.4$, $b=0.01$, $m=0.6$, $\alpha=74.1$, $t_s=13.8$, $\mathcal{R}_0=0.001$).} \label{FIG6}
\end{figure}

The energy conditions for BR model can be derived as,
\begin{eqnarray}
\rho+p&=& -\frac{a(6)^{m-1}m(1-2m)\left(\frac{\alpha}{t_s-t}\right)^{2m-2}\frac{\alpha}{(t_s-t)^2}}{4\pi}\left[1-\kappa_1\right], \nonumber \\
\rho+3p&=&-\frac{a (6)^{m-1} (1-2m) \left(\frac{\alpha}{t_s-t}\right)^{2m-2}} {4\pi(1+2\kappa)}\left[3\left(\frac{\alpha}{t_s-t}\right)^2+m (3+3\kappa-\kappa_1-3\kappa \kappa_1)\frac{\alpha}{(t_s-t)^2}\right], \nonumber \\
\rho-p&=&\frac{a (6)^{m-1} (1-2m)  \left(\frac{\alpha}{t_s-t}\right)^{2m-2}} {4\pi(1+2\kappa)}\left[3\left(\frac{\alpha}{t_s-t}\right)^2+m (1-\kappa+\kappa_1+\kappa \kappa_1)\frac{\alpha}{(t_s-t)^2}\right].\label{eq.26}
\end{eqnarray}
The energy conditions are shown in FIG. \ref{FIG6} (left panel), where we observe that the NEC remains just below the null line and no change are seen in the entire evolution process from early to late time. The SEC remains negative and thus violates. It decreases marginally from early to late time and at sufficient large time, it suddenly decreases. The DEC appears as a mirror image of SEC, it shows a marginal increment and at sufficient late time increases abruptly. 
The geometric parameters can be obtained as,
\begin{eqnarray}
j&=&\frac{(\alpha+1)(\alpha +2)\left(1+R_{0}(t_{s}-t)^{\alpha}\right)^2}{\alpha^2},\nonumber\\
s&=&\frac{(\alpha+1)(\alpha +2)(\alpha +3)\left(1+R_{0}(t_{s}-t)^{\alpha}\right)^3}{\alpha^3},\nonumber\\
l&=&\frac{(\alpha+1)(\alpha +2)(\alpha +3)((\alpha +4))\left(1+R_{0}(t_{s}-t)^{\alpha}\right)^4}{\alpha^4}.\label{eq.27}
\end{eqnarray}
The state finder pair and the lerk parameter are presented in FIG. \ref{FIG6} (right panel). All these geometrical parameters gradually decreases from higher positive value to lower value. At the late time, all are approaching to 1, however the jerk parameter approaches early as compared to the other parameters. The evolution of jerk, snap and lerk parameter starts from lower to higher initial value. Since both the jerk and snap parameter are approaching to 1, indicating a gradual evolution towards $SCDM$ behaviour.    
\subsection{The PR Model}
The cosmos begins in the infinite past from a phase where the scale factor was zero, but the Hubble parameter was a constant. This situation is known as the early phase Pseudo-bang as the characteristics of this are similar to the fate of PR. Another set of dark energy models, where the energy density bounded from above but increases monotonically leads to dissolute the bound structure and termed as the PR cosmological model. We consider here the scale factor of PR model as,
\begin{equation}
\mathcal{R}=R_1 ~~exp\left[H_0 t+H_1\frac{1}{\eta e^{\eta t}}\right],  \label{eq.28}
\end{equation}
where $R_1$ be the integrating constant, $H_0$, $H_1$ and $\eta$ are parameters of the scale factor. We find the Hubble parameter for the scale factor \eqref{eq.28} to be, $H=H_0-\frac{H_1}{e^{\eta t}} $. The Hubble parameter remains finite at infinite time and to obtain the negative deceleration parameter the scale factor parameter $H_1$, $\eta$ to be positive. The deceleration parameter becomes, $q=-1-\frac{\eta H_{1}e^{-\eta t}}{(H_{0}-H_{1}e^{-\eta t})^2}$.
\begin{figure} [H]
\centering
\includegraphics[width=85mm]{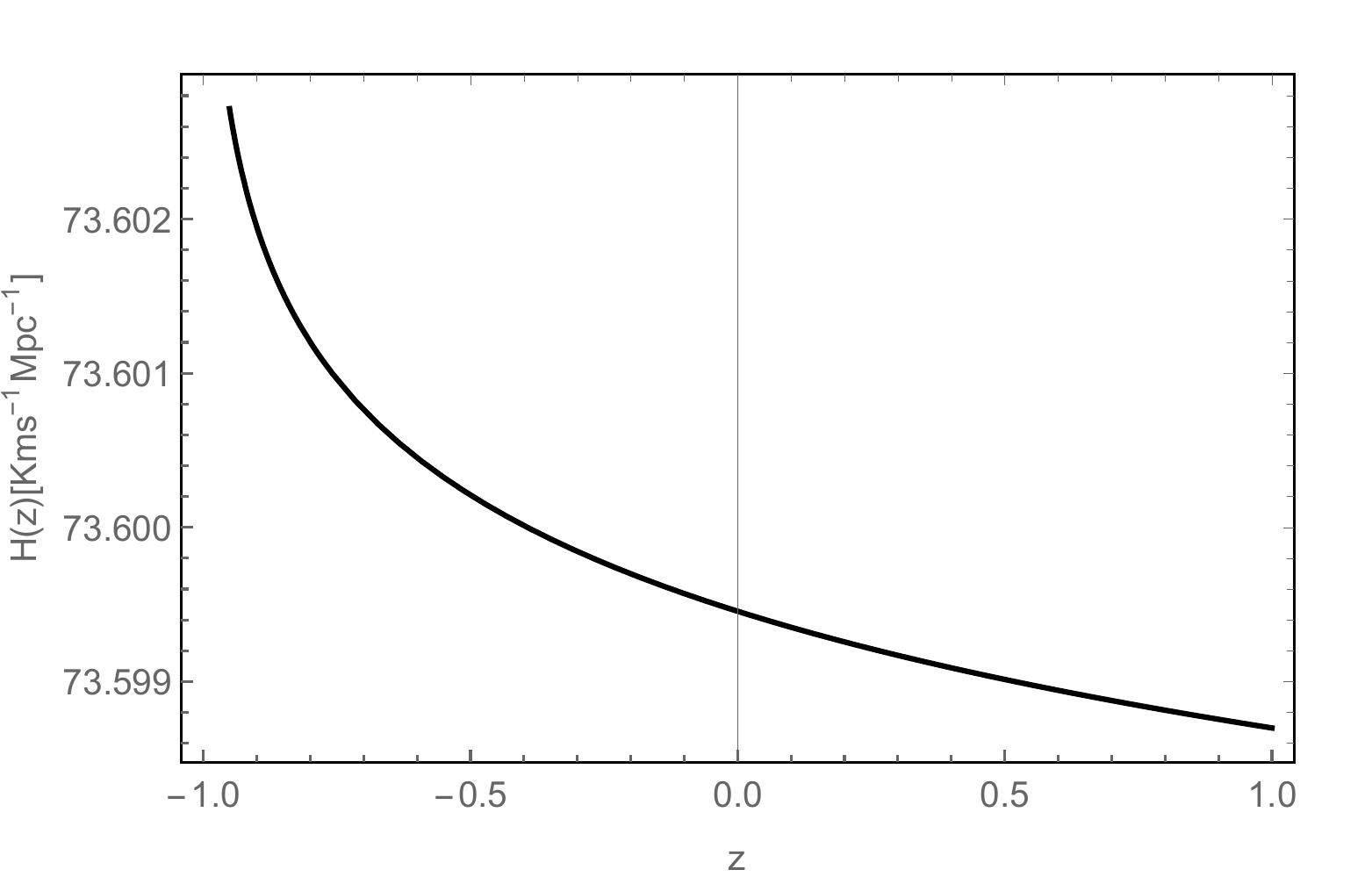}
\includegraphics[width=85mm]{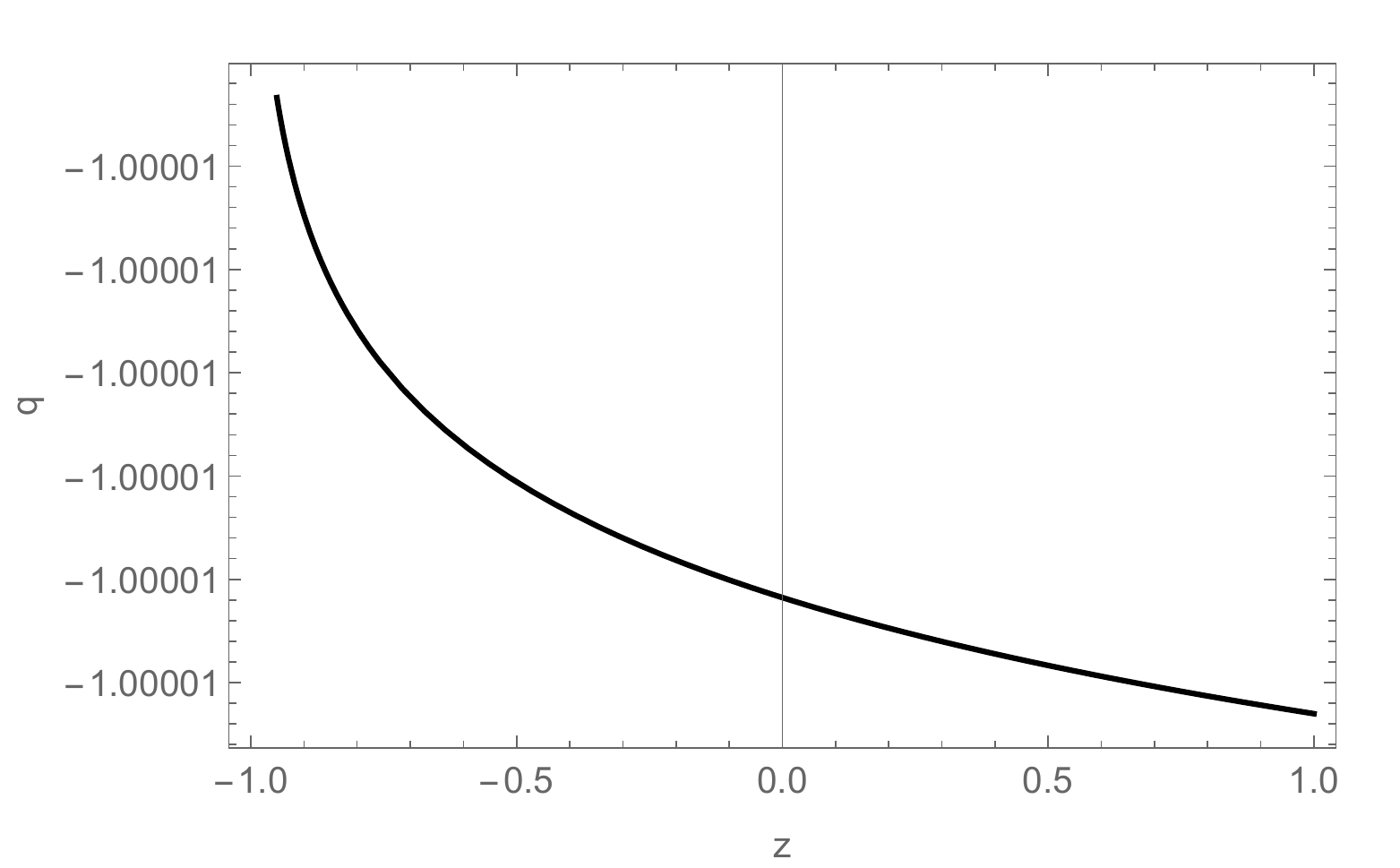}
\caption{Behaviour of Hubble parameter (left panel) and deceleration parameter (right panel) in redshift, ($H_0=73.8$, $H_1=0.2$, $\eta=0.4$).} \label{FIG7}
\end{figure}

The Hubble and deceleration parameters are shown in FIG. \ref{FIG7}. Both the parameters increase gradually respectively in the positive and negative domain. The present value of the Hubble parameter is obtained from the model as $\approx 73.60 Kms^{-1}Mpc^{-1}$  and the deceleration parameter as  $q\approx -1.000012$ which are much closer to the recent observational predictions.
Now the dynamical parameters  can be obtained as,
\begin{eqnarray}
p&=& -\frac{a(6)^{m-1}(1-2m) \left(H_0-H_1 e^{-\eta t}\right)^{2m-2}}{8\pi(1+2\kappa)}\left[3\left(H_0-H_1 e^{-\eta t}\right)^2+m(2+\kappa-\kappa \kappa_1)(\eta H_1 e^{-\eta t})\right], \label{eq.29}\\
\rho&=&\frac{a(6)^{m-1}(1-2m)\left(H_0-H_1 e^{-\eta t}\right)^{2m-2}}{8\pi(1+2\kappa)}\left[3\left(H_0-H_1 e^{-\eta t}\right)^2-m(3\kappa-2\kappa_1-3\kappa \kappa_1)(\eta H_1 e^{-\eta t})\right], \label{eq.30}\\
\omega &=& -1-\frac{2m (1-\kappa_1+2\kappa-2\kappa \kappa_1)(\eta H_1 e^{-\eta t})}{3\left(H_0-H_1 e^{-\eta t}\right)^2-m(3\kappa-2\kappa_1-3\kappa \kappa_1)(\eta H_1 e^{-\eta t})}.\label{eq.31}
\end{eqnarray}
\begin{figure} [H]
\centering
\includegraphics[width=85mm]{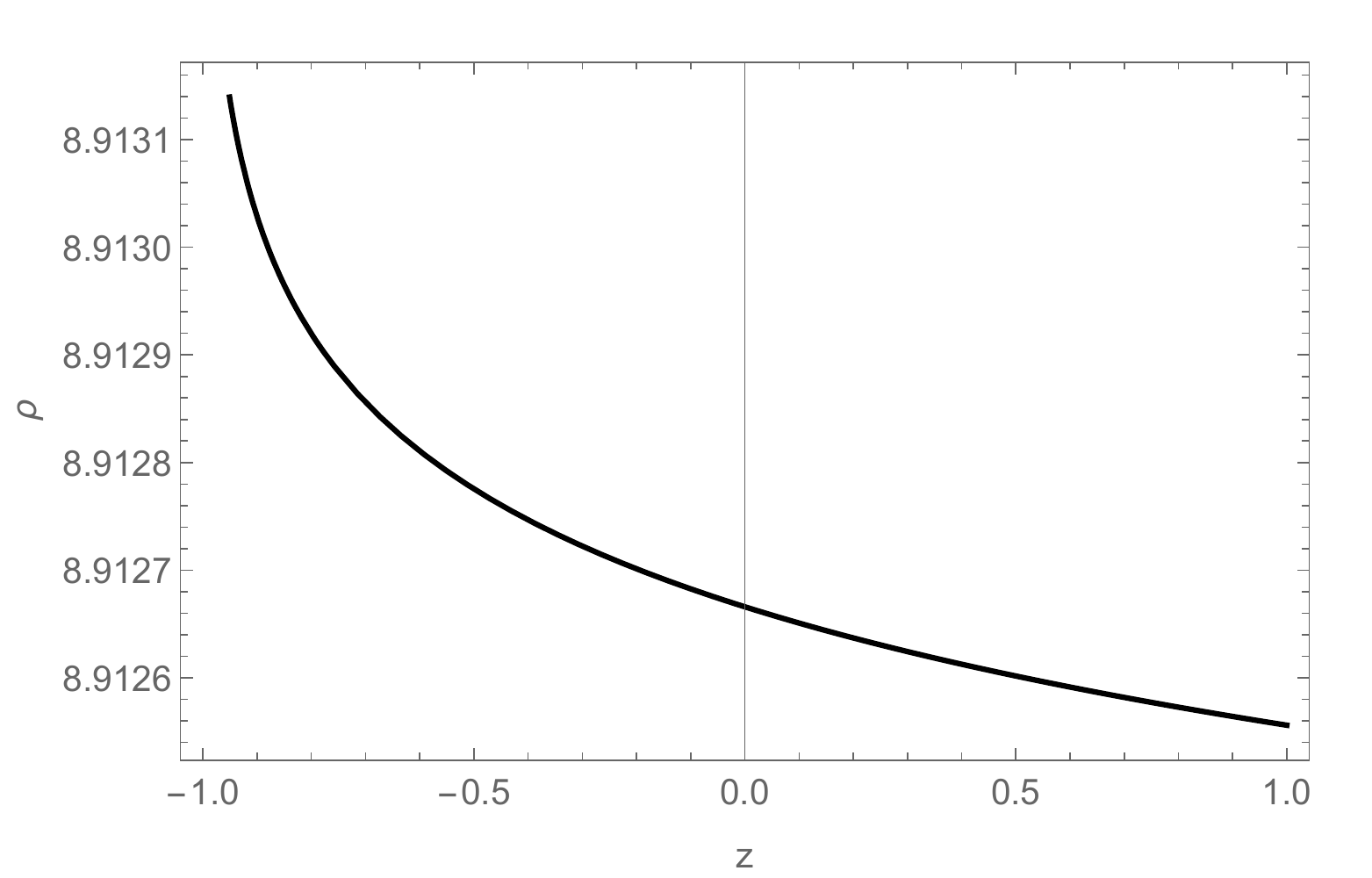}
\includegraphics[width=85mm]{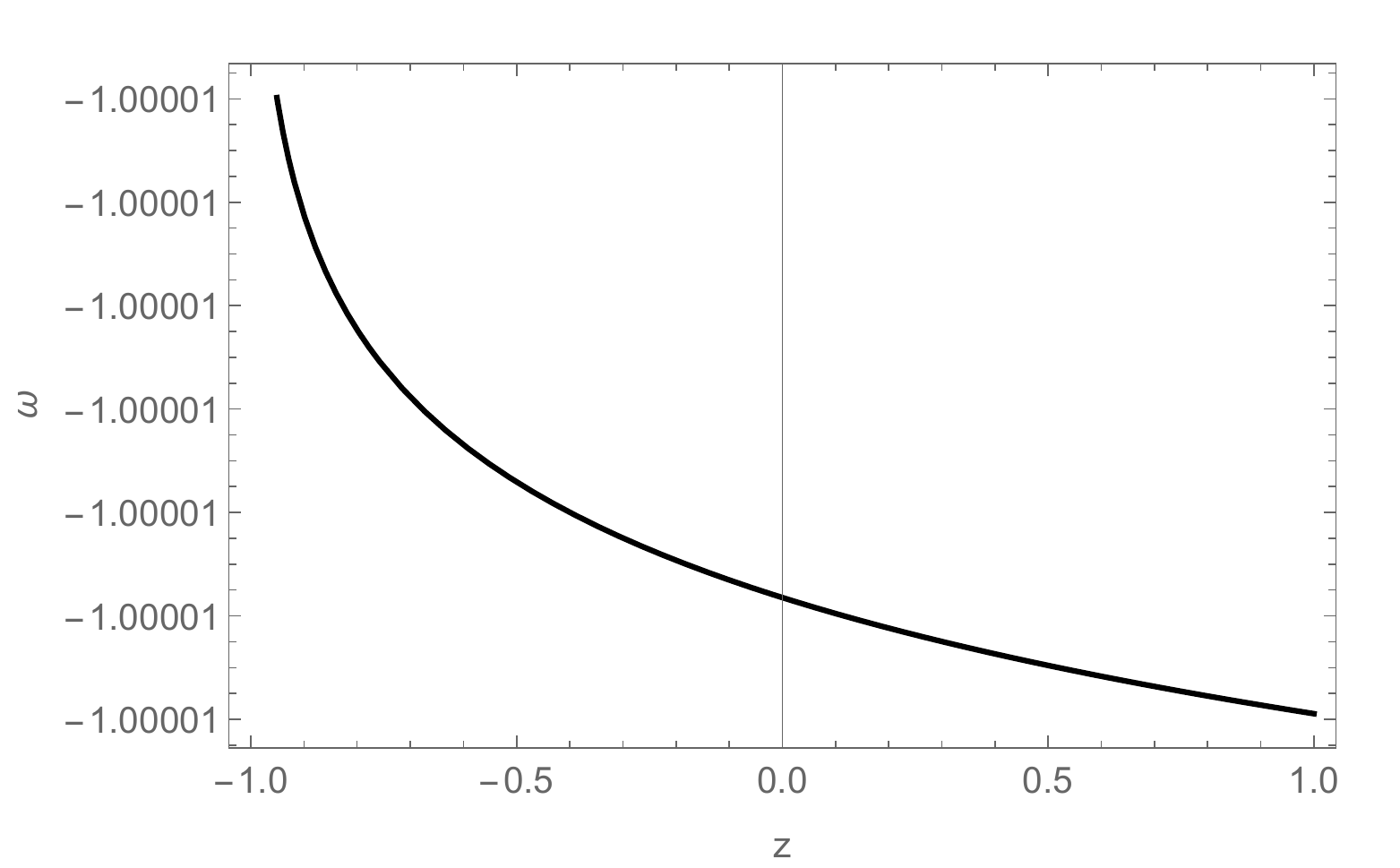}
\caption{Behaviour of energy density (left panel) and EoS parameter (right panel) in redshift, ($a=-4.4$, $b=0.01$, $m=0.6$, $H_0=73.8$, $H_1=0.2$, $\eta=0.4$).} \label{FIG8}
\end{figure}
The energy density becomes positive during the entire evolution and it is increasing from early to late times [FIG.  \ref{FIG8} (left panel)]. The present value of the EoS parameter obtained to be very close to $-1$, so the model is aligning with the concordant $\Lambda$CDM model [FIG.  \ref{FIG8} (right panel)]. The finite time future singularity could not be noticed, thereby it avoids the PR singularity. So, the modification of geometry in the action enabling the model to avoid any kind of PR singularity. 
The energy conditions of the PR model are,

\begin{eqnarray}
\rho+p&=& -\frac{a(6)^{m-1}(1-2m)\left(H_0-H_1 e^{-\eta t}\right)^{2m-2}}{4\pi}\left[m(1-\kappa_1)(\eta H_1e^{-eta t})\right], \nonumber \\
\rho+3p&=&-\frac{a (6)^{m-1} (1-2m) \left(H_0-H_1 e^{-\eta t}\right)^{2m-2}} {4\pi(1+2 \kappa)}\left[3\left(H_0-H_1 e^{-\eta t}\right)^2+m (3+3\kappa-\kappa_1-3\kappa \kappa_1)(\eta H_1 e^{-\eta t})\right], \nonumber \\
\rho-p&=&\frac{a (6)^{m-1} (1-2m)  \left(H_0-H_1 e^{-\eta t}\right)^{2m-2}} {4\pi(1+2\kappa)}\left[3\left(H_0-H_1 e^{-\eta t}\right)^2+m (1-\kappa+\kappa_1+\kappa \kappa_1)(\eta H_1 e^{-\eta t})\right].\label{eq.32}
\end{eqnarray}

\begin{figure} [H]
\centering
\includegraphics[width=85mm]{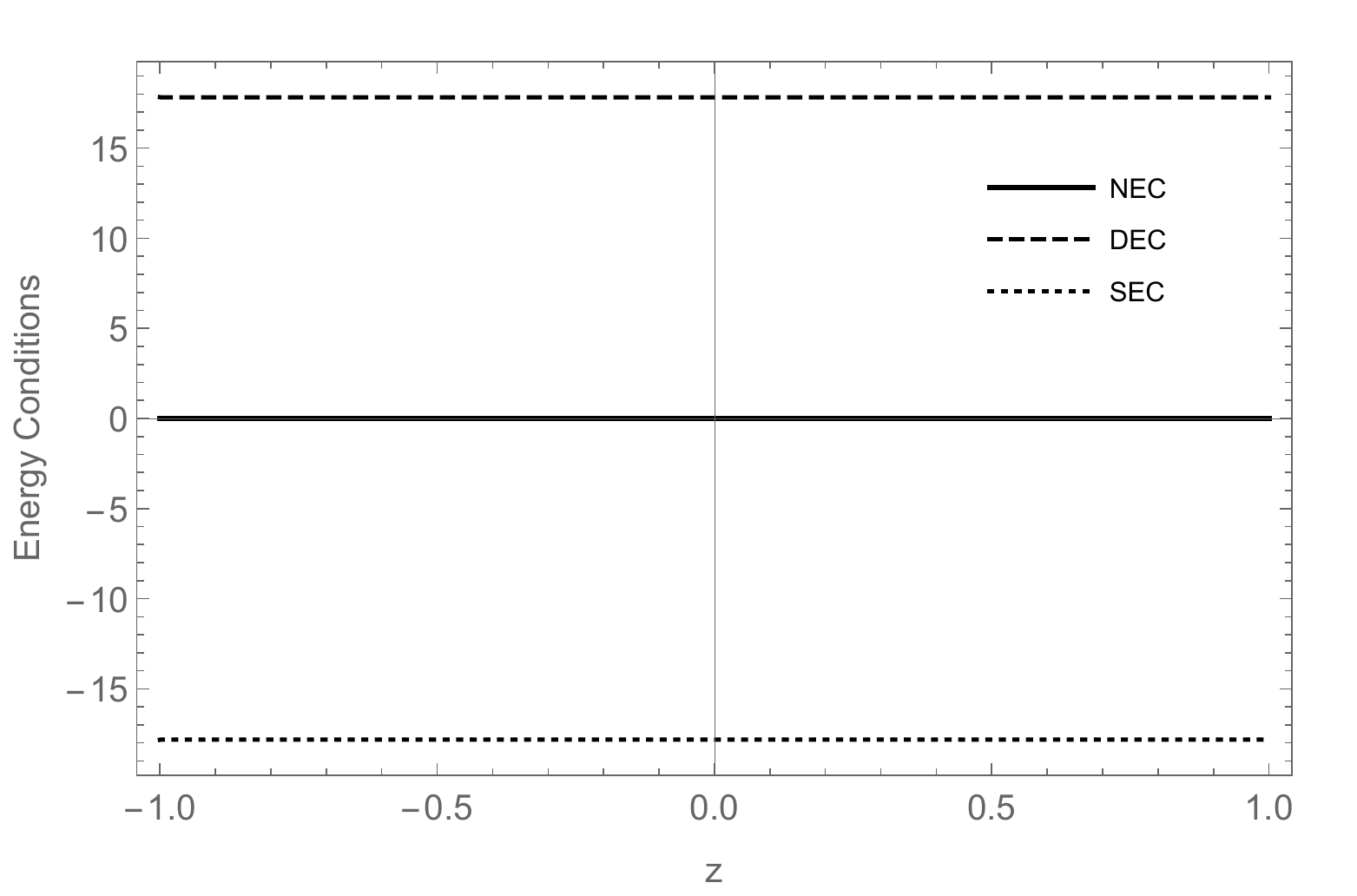}
\includegraphics[width=85mm]{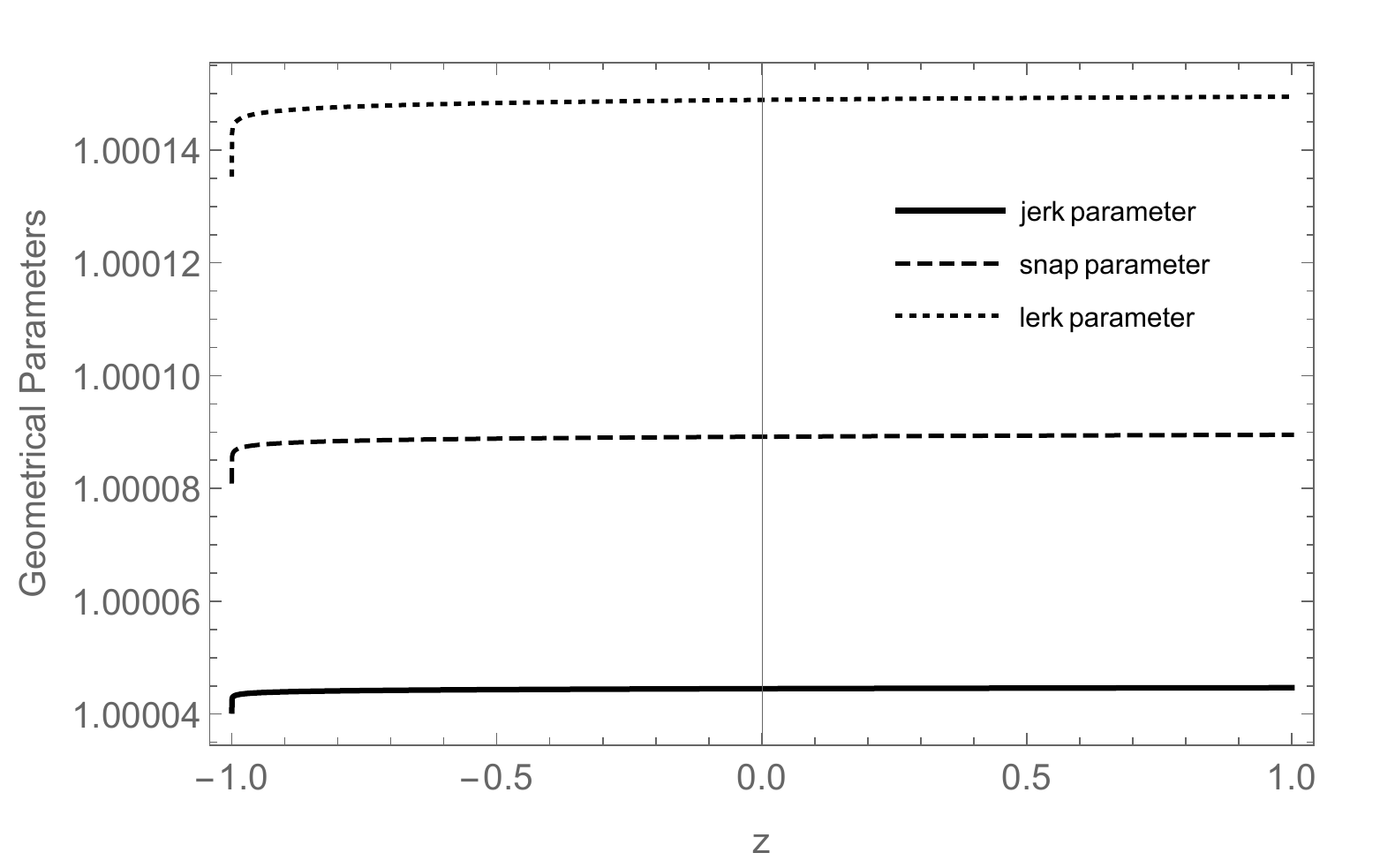}
\caption{Behaviour of energy conditions (left panel) and geometrical parameters (right panel) in redshift, ($a=-4.4$, $b=0.01$, $m=0.6$, $H_0=73.8$, $H_1=0.2$, $\eta=0.4$).} \label{FIG9}
\end{figure}

All the energy conditions almost remain constant throughout. The NEC traverses close to the null line, the DEC remains positive and SEC entirely in the negative domain. Thus the violation of SEC and satisfaction of DEC has been observed in the model. This behaviour supports the accelerating behaviour of the model [FIG.  \ref{FIG9} (right panel)].  
The geometrical parameters of PR models are, 
\begin{eqnarray}
j&=&1-\frac{\eta  e^{-2\eta t}H_{1}  \left(e^{\eta t} (\eta -3H_{0})+3H_{1}\right)}{\left(H_{0} -H_{1} e^{-\eta t} \right)^3}, \nonumber\\
s&=&\frac{e^{-3\eta t}\left[H_{0}^{4} e^{3 \eta t}+e^{ \eta t}H_{1}^{2} \left(6 H_{0}^{2}-12 H_{0} \eta +7 \eta ^2\right)-e^{2 \eta t}H_{1}(2H_{0}-\eta ) \left(2 H_{0}^{2}-2H_{0} \eta +\eta ^2\right)-2 H_{1}^{3} (2H_{0}-3 \eta )+e^{-\eta t}H_{1}^{4}\right]}{\left(H_{0} -H_{1} e^{-\eta t}\right)^4},\nonumber\\
l&=&\frac{e^{-4\eta t}\left[H_{0}^{5} e^{4 \eta t}-5 e^{ \eta t}H_{1}^{3} \left(2 H_{0}^{2}-6 H_{0} \eta +5 \eta ^2\right)+5 e^{2 \eta t}H_{1}^{2} (H_{0}-\eta ) \left(2 H_{0}^{2}-4 H_{0} \eta +3 \eta ^2\right)\right]}{\left(H_{0} -H_{1} e^{-\eta t}\right)^5},\nonumber\\
&-&\frac{e^{-4\eta t}\left[e^{3 \eta t} H_{1}\left(5 H_{0}^{4}-10 H_{0}^{3} \eta +10 H_{0}^{2} \eta ^2-5 H_{0} \eta ^3+\eta ^4\right)-5 H_{1}^{4} (H_{0}-2 \eta )+e^{-\eta t} H_{1}^{5}\right]}{\left(H_{0} -H_{1} e^{-\eta t}\right)^5}. \label{eq.33}
\end{eqnarray}

The three geometric parameters decrease very slowly from early to late Universe and the snap parameter appears in between the other two parameters. The lerk parameter evolves from a higher value as compared to snap and jerk parameter. Though all the parameters are not almost equal to 1, there is very small deviations. One interesting thing noticed is that all the three parameters experience a sudden dip during infinite future. The responsibility of this behaviour goes to the choice of the parameters in order to keep the dynamical parameters in the desired range.
\section{Results and Conclusion}
The late time cosmic acceleration issue could not be well addressed in the context of GR, however the geometrically modified gravity could give some insight into this issue of modern cosmology. In addition, because of the acceleration, another challenge is whether the Universe would torn apart in finite or infinite future. In an attempt to address this issue, in this paper, we have presented three models related to the rip singularity in the extended symmetric teleparallel gravity. Initially, we have adjusted the scale factor parameters in such a manner that the Hubble parameter shows the present value as suggested by cosmological observations. Accordingly appropriate behaviour of the deceleration parameter could be obtained. The avoidance of a finite time future singularity becomes a central issue in cosmological studies. In this context, the role of the non-metricity in place of the usual Ricci scalar  in avoiding of the finite time future singularity may not be ruled out.\\

The EoS parameter in case of LR model though shows a phantom type behaviour but remains very close to the $\Lambda$CDM line whereas in BR model it entirely remains on the $\Lambda$CDM line. At the same time in PR model, it shows similar behaviour as in LR model except the fact that in PR model it remains in a further narrow range. As required in the modified theories of gravity, here also in all three models violation of SEC and satisfaction of DEC are obtained. The interesting behaviour we noticed is that the NEC appears just below the null line. It indicates that the contribution from NEC is almost negligible in these models. Apart from the deceleration parameter, the jerk, snap and lerk parameters are also analysed for these models. Though there are certain physical behaviours can be interpreted from the value of $(j,s)$ pair, but no observational evidences are available on lerk parameter. So, in case of both LR and PR model these parameters decrease marginally over the time and experience sudden deep and some late time, but in BR model the SEC and DEC respectively shows the dip and peak. In conclusion we can interpret that no singularity scenario appear in the accelerating models, so this study in $f(Q,T)$ gravity may give new insight into resolving the singularity issue.     

\section*{Acknowledgement}
LP acknowledges Department of Science and Technology (DST), Govt. of India, New Delhi for awarding INSPIRE fellowship (File No. DST/INSPIRE Fellowship/2019/IF190600) to carry out the research work. SAK acknowledges the financial support provided by University Grants Commission (UGC) through Junior Research Fellowship (UGC Ref. No.: 191620205335), to carry out the research work. BM and SKT thank IUCAA, Pune, India for supporting through the visiting associateship program.


\begin{thebibliography}{99}
\section*{References}

\bibitem{Riess98} A.G. Riess, et al., \href{https://doi.org/10.1086/300499}{Astron. J., 116 (1998) 1009}.

\bibitem{Perlmutter99} S. Perlmutter, et al., \href{https://doi.org/10.1086/307221}{Astrophys. J. 517 (1999) 565}.

\bibitem{Ade16} P.A.R. Ade, et al., Planck collaboration, Planck 2015 results. XIII. Cosmological parameters, \href{https://doi.org/10.1051/0004-6361/201525830}{Astron. Astrophys. 594 (2016) A13}.

\bibitem{Aghanim20} N. Aghanim, et al., Planck 2018 results, VI. Cosmological parameters, \href{https://doi.org/10.1051/0004-6361/201833886}{Astron. Astrophys. 641 (2020) A6}.

\bibitem{Caldwell03} R. R. Caldwell, M. Kamionkowski, N. N. Weinberg, \href{https://doi.org/10.1103/PhysRevLett.91.071301} {Phys. Rev. Lett. 91 (2003) 071301}.

\bibitem{Frampton03} P. H. Frampton, T. Takahashi, \href{https://doi.org/10.1016/S0370-2693(03)00208-9}{Phys. Lett. B 557 (2003) 135}.

\bibitem{Nesseris04} S. Nesseris, L. Perivolaropoulos, \href{https://doi.org/10.1103/PhysRevD.70.043531}{Phys. Rev. D 70 (2004) 043531}.

\bibitem{Scherrer05} R. J. Scherrer, \href{https://doi.org/10.1103/PhysRevD.71.063519}{Phys. Rev. D 71 (2005) 063519}.

\bibitem{Valentino16} E.D. Valentino, A. Melchiorri and J. Silk, \href{https://doi.org/10.1016/j.physletb.2016.08.043}{Phys. Lett. B 761 (2016)242}.

\bibitem{Vagnozzi20} S. Vagnozzi, \href{https://doi.org/10.1103/PhysRevD.102.023518} {Phys. Rev. D 102 (2020) 023518}.

\bibitem{Valentino21}E.D. Valentino, A.Mukherjee, A.A. Sen, \href{https://doi.org/10.3390/e23040404} {Entropy 23 (2021) 404}.



\bibitem{Sahni03} V. Sahni, Y. Shtanov, \href{https://doi.org/10.1088/1475-7516/2003/11/014}{JCAP 0311 (2003) 014}; H. Stefancic, \href{https://doi.org/10.1103/PhysRevD.71.084024}{Phys. Rev. D 71 (2005) 084024}; A. Yurov, \href{https://doi.org/10.1140/epjp/i2011-11132-7}{Eur. Phys. J. Plus 126 (2011) 132}; P.H. Frampton, K.J. Ludwick, R.J. Scherrer, \href{https://doi.org/10.1103/PhysRevD.84.063003}{Phys. Rev. D 84, (2011) 063003}. 

\bibitem{Nojiri05} S. Nojiri, S. D. Odintsov, S. Tsujikawa, \href{https://doi.org/10.1103/PhysRevD.71.063004}{Phys.Rev.D 71 (2005) 063004}.

\bibitem{Dabrowski03} M.P. Dabrowski, T. Stachowiak and M. Szydlowski, \href{https://arxiv.org/pdf/hep-th/0307128} {Phys.Rev.D 68 (2003) 103519}.


\bibitem{Astashenok12} A.V. Astashenok et al., \href{https://doi.org/10.1016/j.physletb.2012.02.039}{Phys. Lett. B 709 (2012) 396}. 

\bibitem{Starobinsky00} A.A. Starobinsky, \href{https://www.worldscientific.com/doi/abs/10.1142/9789812793324_0008}{Grav. Cosmol. 6 (2000) 157}; S. Nojiri, S.D. Odintsov, \href{https://doi.org/10.1016/S0370-2693(03)00594-X}{Phys. Lett. B 562 (2003) 147}; P.F. Gonzalez-Diaz, \href{https://doi.org/10.1103/PhysRevD.69.063522}{Phys. Rev. D 69 (2004) 063522}; S. Nojiri, S.D. Odintsov, \href{https://doi.org/10.1103/PhysRevD.70.103522}{Phys. Rev. D 70 (2004) 103522}; V. Faraoni, \href{https://doi.org/10.1088/0264-9381/22/16/008}{Class. Quant. Grav. 22 (2005) 3235}; B. McInnes, \href{https://doi.org/10.1016/j.nuclphysb.2005.04.025}{Nucl. Phys. B 718 (2005) 55}. 



\bibitem{Frampton12} P.H. Frampton, K.J. Ludwick, R.J. Scherrer, \href{https://doi.org/10.1103/PhysRevD.85.083001} {Phys.Rev.D 85 (2012) 083001}.

\bibitem{Frampton11} P. H. Frampton, K. J. Ludwick, R. Scherrer, \href{https://doi.org/10.1103/PhysRevD.84.063003} {Phys. Rev. D 84, (2011) 063003}. 


\bibitem{Gomez13} D. Saez-Gomez, \href{https://doi.org/10.1088/0264-9381/30/9/095008}{Class. Quantum Grav. 30 (2013) 095008}.

\bibitem{Makarenko13} A.N. Makarenko, V.V. Obukhov, I. V. Kirnos, \href{https://doi.org/10.1007/s10509-012-1240-1}{Astrophys. Space Sci. 343 (2013) 481}.

\bibitem{Houndjo14} M.J.S. Houndjo et al., \href{https://arxiv.org/abs/1207.1646}{Eur. Phys. J. Plus 129 (2014) 171}.

\bibitem{Mishra20} B. Mishra, S.K. Tripathy, \href{https://doi.org/10.1088/1402-4896/abb0ab}{Phys. Scr. 95 (2020) 095004}.

\bibitem{Vasilev19} T.V. Vasilev, M. Bouhmadi-Lopez, P. Martin-Moruno, \href{https://doi.org/10.1103/PhysRevD.100.084016}{Phys. Rev. D 100 (2019) 084016}. 

\bibitem{Vasilev21} T.V. Vasilev, M. Bouhmadi-Lopez, P. Martin-Moruno, \href{https://doi.org/10.1103/PhysRevD.103.124049}{Phys. Rev. D 103 (2021) 124049}. 

\bibitem{Bakry21} M. A. Bakry, A. T. Shafeek, \href{https://doi.org/10.1134/S0202289321010047}{Grav. Cosmol. 27 (2021) 89}.

\bibitem{Ray21} P. Ray et al., \href{https://doi.org/10.1002/prop.202100086}{Fortschr. Phys. (2021) 2100086}.

\bibitem{Hanafy20} W. El Hanafy, E. N. Saridakis, \href{https://doi.org/10.1088/1475-7516/2021/09/019}{Journal of Cosmology and Astroparticle Physics 09 (2021) 019}.

\bibitem{Xu19} Y. Xu, G. Li,T. Harko, S. Liang, \href{https://doi.org/10.1140/epjc/s10052-019-7207-4}{\textit{Eur. Phys. J. C}, \textbf{79}  (2019) 708}.

\bibitem{Nester99} J.M. Nester, H.-J. Yo, \href{https://arxiv.org/abs/gr-qc/9809049v2}{Chin. J. Phys. 37 (1999) 113}.

\bibitem{Jimenez18} J. Beltran Jimenez, et al., \href{https://doi.org/10.1103/PhysRevD.98.044048}{Phys. Rev. D 98 (2018) 044048}.

\bibitem{Jimenez19} J. Beltran Jimenez, L. Heisenberg, T.S. Koivisto, \href{https://doi.org/10.3390/universe5070173} {Universe 5 (2019) 173}.

\bibitem{Pati21} L. Pati, B.Mishra, S.K. Tripathy, \href{https://doi.org/10.1088/1402-4896/ac0f92}{Phys. Scr. 96 (2021) 105003}.

\bibitem{Agrawal21} A.S. Agrawal et al.,\href{https://doi.org/10.1016/j.dark.2021.100863}{Phys. Dark Univ.33 (2021) 100863}.

\bibitem{Zia21} R. Zia, D. C. Maurya, A. K. Shukla, \href{https://ui.adsabs.harvard.edu/link_gateway/2021IJGMM..1850051Z/doi:10.1142/S0219887821500511}{Int. J. Geom. Methods Mod. Phys. 18 (2021) 2150051}.

\bibitem{Godani21} N. Godani, G.C. Samanta, \href{https://doi.org/10.1142/S0219887821501930}{Int. J. Geom. Methods Mod. Phys. 18 (2021)}.

\bibitem{Najera21} A. Najera, A. Fajardo,\href{https://arxiv.org/abs/2104.14065v2}{arXiv:2104.14065 (2021)}.

\bibitem{Camarena20} D Camarena, V. Marra, \href{https://doi.org/10.1103/PhysRevResearch.2.013028}{Phys. Rev. Research. 2 (2020) 013028}.

\bibitem{Hawking99} S.W. Hawking, G.F.R. Ellis, The Large Structure of Space-Time, \href{https://scholar.google.com/scholar?cites=4686600064207465389&as_sdt=2005&sciodt=0,5&hl=en}{Cambridge University Press, Cambridge, (1999)}.

\end{thebibliography}
\end{document}